\documentclass[11pt]{article}%
\usepackage{amsfonts}
\usepackage{xcolor}
\colorlet{linkequation}{green}
\usepackage{amsmath}%
\usepackage{amssymb}%
\usepackage{graphicx}
\usepackage{float}
\usepackage{longtable}
\usepackage[utf8]{inputenc}
\usepackage{hyperref}
\usepackage{natbib} 
\usepackage{mathdots}
\usepackage{upgreek}
\usepackage{amsmath}
\usepackage{algorithmicx}
\usepackage[ruled,vlined]{algorithm2e}
\SetAlgoCaptionSeparator{ }

\usepackage{enumerate}

\newcommand{\fref}[1]{Fig.~\ref{#1}}

\newcommand{\tref}[1]{Table~\ref{#1}}
\newcommand{\sref}[1]{Section~\ref{#1}}

\newcommand*{\cref}[1]{%
  \begingroup
    \hypersetup{
      linkcolor=linkequation,
      linkbordercolor=linkequation,
    }%
    \ref{#1}%
  \endgroup
}

\newenvironment{method}[1][htb]
  {
   \begin{algorithm}[#1]%
  }{\end{algorithm}}

\providecommand{\U}[1]{\protect\rule{.1in}{.1in}}
\begin{document}

\title{Entanglement-Gradient Routing for Quantum Networks}
\author{Laszlo Gyongyosi$^{1,2,3,}$\thanks{E-mail: \href{mailto:l.gyongyosi@soton.ac.uk}{l.gyongyosi@soton.ac.uk}} , Sandor Imre$^{2}$\\
$^{1}$School of Electronics and Computer Science\\University of Southampton\\Southampton SO17 1BJ, UK\\
$^{2}$Department of Networked Systems and Services\\Budapest University of Technology and Economics\\Budapest, H-1117 Hungary\\
$^{3}$MTA-BME Information Systems Research Group\\Hungarian Academy of Sciences\\Budapest, H-1051 Hungary}
\date{}

\maketitle
\begin{abstract}
We define the entanglement-gradient routing scheme for quantum repeater networks. The routing framework fuses the fundamentals of swarm intelligence and quantum Shannon theory. Swarm intelligence provides nature-inspired solutions for problem solving. Motivated by models of social insect behavior, the routing is performed using parallel threads to determine the shortest path via the entanglement gradient coefficient, which describes the feasibility of the entangled links and paths of the network. The routing metrics are derived from the characteristics of entanglement transmission and relevant measures of entanglement distribution in quantum networks. The method allows a moderate complexity decentralized routing in quantum repeater networks. The results can be applied in experimental quantum networking, future quantum Internet, and long-distance quantum communications.
\end{abstract}
%{\small {\bf Keywords:} quantum networking; quantum repeater; quantum entanglement; quantum communication; quantum Shannon theory.}
%\label{keywords}

\section{Introduction}
\label{sec1}
 Finding the shortest path in an entangled quantum network is desired for improving the efficiency of quantum repeater networks of the quantum Internet, and of long-distance quantum communications [\cref{ref1}-\cref{ref11}], [\cref{ref25}-\cref{ref26}], [\cref{ref28}-\cref{ref29}]. By definition, in an entangled quantum network, the quantum nodes share quantum entanglement. A transmitter and receiver node is separated by several intermediate quantum repeaters, and a chain of entangled links forms a path (entangled path) between the source and destination [\cref{ref30}-\cref{ref47}]. The level of an entangled link between the quantum nodes determines the achievable hop distance and the number of spanned intermediate nodes. Since quantum networks integrate different levels of entangled links, a shortest path between a source and destination quantum node has to be found in a multi-level quantum network architecture [\cref{ref1}-\cref{ref11}], [\cref{ref30}-\cref{ref47}]. An entangled link has several relevant attributes, such as the level of entanglement (number of nodes spanned by a source-destination path), the entanglement throughput of the link that quantifies the number of entangled states transmitted at a particular fidelity [\cref{ref1}-\cref{ref4}]. The quantum nodes receive and store the entangled states in their local quantum memories [\cref{ref33}-\cref{ref48}] for further extension of the range of entanglement. In the quantum nodes, the number of incoming entangled states represents a crucial parameter from the modeling perspective, along with the mean number of received states (observation rate), and with the reduction in the amount of received entangled states (decay rate). 

In this work, we define the \textit{entanglement-gradient routing} scheme for quantum repeater networks. The proposed routing framework fuses the fundamentals of swarm intelligence [\cref{ref12}-\cref{ref17}] and the results of quantum Shannon theory. Swarm intelligence provides nature-inspired solutions for problem solving. In general, it refers to some population-based meta-heuristics that are motivated by the behavior of living entities (ant colony, bee colony, flock of birds, particle swarm, bacteria foraging, etc.) interacting locally both with each other and the environment. Swarm intelligence has a wide range of applications in real-world problems, ranging from optimization tasks, data mining, computer science, database searching and knowledge discovery to bioinformatics and social networks.

Our entanglement-gradient routing scheme uses finds the shortest path in a decentralized manner. Motivated by the models of social insect behavior, the routing is relying on several parallel threads, where the threads represent simple, locally interacting individual swarms. 

The routing and path selection for quantum repeater networks has been studied in several different works [\cref{ref2}-\cref{ref5}]. Without loss of generality, most of these approaches utilized a variance of the well-known Dijsktra's algorithm [\cref{ref24}] for the determination of the shortest path in the quantum network [\cref{ref1}-\cref{ref5}]. On the other hand, these works have successfully confirmed that a shortest path algorithm from the traditional context is implementable and works well in a quantum environment. In our work we step further, and inject significant novelties to the procedures of routing and path selection in quantum networks. Our framework breaks with the practice of implementing a Dijsktra-variant algorithm or other, well-known traditional routing protocol in a quantum environment [\cref{ref1}]. In our solution, the shortest paths are determined by a biologically-inspired, decentralized algorithm that takes into account the physical-layer attributes of the entanglement establishment and the quantum transmission. 

The \textit{entanglement gradient} coefficient quantifies the attractiveness of entangled links and paths for the threads in the quantum repeater network. Each thread acts in a localized manner and the threads are attracted by the entanglement gradients of the paths. The routing is based on metrics that use the tools of quantum Shannon theory. The metrics are derived from the characteristics of entanglement transmission and relevant physical and statistical measures of entanglement distribution. To measure the relevance of a particular entangled link, we define the \textit{entanglement utility} coefficient. Using the entanglement throughput characteristic extractable from the quantum network, we define the \textit{link entanglement gradient} coefficient. We then extend the entanglement gradient for entangled paths (\textit{path entanglement gradient} coefficient), which refers to a path formulated by entangled links. 

The aim of using the threads is to find the most attractive path in the quantum network with a highest entanglement gradient (i.e., lowest inverse entanglement gradient) similar to the methods of swarm intelligence. The entanglement gradient evolves in time, decaying as the entanglement throughput deviates from a mean value (decay rate coefficient). 

The threads build probabilistic paths between the quantum nodes using simple processing steps to keep minimal the complexity of the scheme. We also include a performance analysis of the routing scheme. The proposed routing method supports a moderate-complexity routing in quantum repeater networks. 

The scheme is straightforwardly applicable by standard physical devices in an experimental quantum networking scenario. A physical implication of a stationary node in our quantum network model can integrate standard photonics devices, quantum memories, optical cavities and other fundamental physical devices [\cref{ref1}, \cref{ref20}-\cref{ref21}]. The quantum transmission between the nodes can be realized via noisy quantum links (e.g., optical fibers, wireless quantum channels, free-space optical channels, etc) and fundamental quantum transmission protocols [\cref{ref21}]. 

Since the method is based on the fundamentals of swarm intelligence theory, the proposed framework allows a fusion with the elements of quantum machine learning [\cref{ref22}-\cref{ref23}]. By utilizing additional functions in the quantum nodes, the model provides a ground for a direct application of a distributed secure quantum machine learning method [\cref{ref27}]. 

The novel contributions of this paper are as follows: 
\begin{itemize}
\item \textit{We provide a nature-inspired, decentralized routing scheme for quantum repeater networks.}
\item \textit{The routing metric utilizes the attributes of entangled links, the properties of entanglement transmission and the statistical distribution of the entangled states in the quantum network.}
\item \textit{The method supports an efficient and moderate-complexity routing in quantum repeater networks by fusing the relevant characteristics of entanglement distribution and swarm intelligence theory.}
\item \textit{The scheme provides an easy experimental implementation by standard photonics devices, provides a useful tool for shortest path finding in quantum Internet and in practical long-distance quantum communications.}
\end{itemize}

This paper is organized as follows. In \sref{sec2}, the preliminaries and definitions are introduced. \sref{sec3} discusses the entanglement gradient of entangled paths, while \sref{sec4} details the entanglement-gradient routing proposed for quantum repeater networks. In \sref{sec5}, a numerical analysis is provided. Finally, \sref{sec6} concludes the paper. Some supplemental information is included in the Appendix.

\section{Preliminaries }
\label{sec2}
In this preliminary section, we summarize the terms and definitions. 
  
\subsection{Entanglement Utility}

 In the proposed model, the relevance of a particular entangled link is characterized by the entanglement utility\textit{ }coefficient, ${\lambda}_{E_{{\mathrm{L}}_l}\left(x,y\right)}$ of an entangled link\textit{ }$E_{L_l}\left(x,y\right)$ between nodes $x$ and $y$, where ${\mathrm{L}}_l$ is the level of the entangled link (By definition, for an ${\mathrm{L}}_l$-level entangled link, the hop distance between quantum nodes $x$ and $y$ is ${\mathrm{2}}^{l\mathrm{-}\mathrm{1}}$).

 This amount is equivalent to the utility of the entangled link $E_{{\mathrm{L}}_l}\left(x,y\right)$ that it has taken in order to arrive at the current node $y$ from $x$ (see \fref{fig1}), and initialized without loss of generality as 
\begin{equation} \label{1)} 
{\lambda }_{E_{{\mathrm{L}}_l}\left(x,y\right)}\mathrm{\ge }\mathrm{0}.                                                                 
\end{equation} 
Let $A$ be the source quantum node and $B$ the target repeater node. Let $y$ be the current node with a direct neighbor $x$ and an established entangled link $E_{{\mathrm{L}}_l}\left(x,y\right)$ between $x$ and $y$ [\cref{ref15}-\cref{ref16}], [\cref{ref18}-\cref{ref19}]. 

 Let $B_F\left(E_{{\mathrm{L}}_l}\left(x,y\right)\right)$ refer to the \textit{entanglement} \textit{throughput} of a given ${\mathrm{L}}_l$-level entangled link $E_{{\mathrm{L}}_l}\left(x,y\right)$ between nodes $\left(x,y\right)$ measured in the number of $d$-dimensional entangled states per sec at a particular entanglement fidelity $F$ [\cref{ref1}], [\cref{ref3}-\cref{ref4}].   

 In our scheme, at a given $B_F\left(E_{{\mathrm{L}}_l}\left(x,y\right)\right)$, the update of an initial ${\lambda }_{E_{{\mathrm{L}}_l}\left(x,y\right)}$ entanglement utility of link $E_{{\mathrm{L}}_l}\left(x,y\right)$ to ${\lambda }'_{E_{{\mathrm{L}}_l}\left(x,y\right)}$ is defined as
\begin{equation} \label{ZEqnNum640853} 
 \begin{array}{l}
{\lambda }'_{E_{{\mathrm{L}}_l}\left(x,y\right)}\\\mathrm{=}{\left(\frac{\mathrm{1}}{{\lambda }_{E_{{\mathrm{L}}_l}\left(x,y\right)}}\mathrm{+}B_F\left(E_{{\mathrm{L}}_l}\left(x,y\right)\right)\right)}^{\mathrm{-}\mathrm{1}}\\\mathrm{=}\frac{{\lambda }_{E_{{\mathrm{L}}_l}\left(x,y\right)}}{\mathrm{1+}B_F\left(E_{{\mathrm{L}}_l}\left(x,y\right)\right){\lambda }_{E_{{\mathrm{L}}_l}\left(x,y\right)}}, \end{array}
\end{equation} 
where $B_F\left(E_{{\mathrm{L}}_l}\left(x,y\right)\right)$ serves as a cost function between node pair $\left(x,y\right)$ which is added to the inverse of the current entanglement utility, i.e., ${\mathrm{1}}/{{\lambda }_{E_{{\mathrm{L}}_l}\left(x,y\right)}}$. 
The update mechanism of \eqref{ZEqnNum640853} is therefore formulates the evolution of entanglement utility in the destination node $y$. Utilizing the fundamental updating methods of swarm intelligence [\cref{ref15}-\cref{ref18}], \eqref{ZEqnNum640853} provides a solution to take into account not just the characteristic of entanglement transmission, but also the physical attributes of quantum links.   
 
\subsection{Link Entanglement Gradient }
The attractiveness of a particular quantum node is characterized by the link entanglement gradient coefficient. Let ${\mathcal{G}}^y_{A,x}$ be the amount of entanglement gradient from source node $A$, on the neighbor node $x$ at $y$, initialized as ${\mathcal{G}}^y_{A,x}\mathrm{\ge }\mathrm{0}$. The entanglement gradient is updated in a particular quantum node $y$, as follows. 

 Motivated by the fundamentals of swarm intelligence theory [\cref{ref15}-\cref{ref19}], using \eqref{ZEqnNum640853} the entanglement gradient ${\mathcal{G}}^y_{A,x}$ at current node $y$ and entangled link $E_{{\mathrm{L}}_l}\left(x,y\right)$ is updated to ${\mathcal{G}}'^y_{A,x}$ as
\begin{equation} \label{ZEqnNum489942} 
{\mathcal{G}}'^y_{A,x}\mathrm{=}{\mathcal{G}}^y_{A,x}f\left(\mathrm{-}\tau \left(\mathit{\Delta}B_F\left(E_{{\mathrm{L}}_l}\left(x,y\right)\right)\right)\right)\mathrm{+}{\lambda }'_{E_{{\mathrm{L}}_l}\left(x,y\right)},                                                         
\end{equation} 
where $\tau \mathrm{\ge }\mathrm{0}$ is a decay rate of entanglement gradient, function $f\left(x\right)$ provides a probability distribution, while the \textit{entanglement throughput deviation} parameter, $\mathit{\Delta}B_F\left(E_{{\mathrm{L}}_l}\left(x,y\right)\right)$, is defined as
\begin{equation} \label{4)} 
\begin{split}
\mathit{\Delta}B_F\left(E_{{\mathrm{L}}_l}\left(x,y\right)\right)\\\mathrm{=}&\left|\frac{\sum^n_{h\mathrm{=1}}{B_F\left(E_{{\mathrm{L}}_l}\left(y,h\right)\right)}}{n}\mathrm{-}B_F\left(E_{{\mathrm{L}}_l}\left(x,y\right)\right)\right|,                                                        
\end{split}
\end{equation} 
where $n$ is the number of direct connections of node $y$, $\sum^n_{h\mathrm{=1}}{B_F\left(E_{{\mathrm{L}}_l}\left(y,h\right)\right)}$ is the total entanglement throughput of all $n$ direct links of node $y$, while $B_F\left(E_{{\mathrm{L}}_l}\left(x,y\right)\right)$ is the entanglement throughput of link $E_{{\mathrm{L}}_l}\left(x,y\right)$ between nodes $y$ and $x$. 

 For all other neighbors $j$,$j\mathrm{=1,\dots ,}n$, $j\mathrm{\in }V\mathrm{-}x$, the entanglement gradient ${\mathcal{G}}^y_{A,i}$ is only decreased by a factor $f\left(\mathrm{-}\tau \left(\mathit{\Delta}B_F\left(E_{{\mathrm{L}}_l}\left(j,y\right)\right)\right)\right)$, thus
\begin{equation} \label{ZEqnNum529081} 
{\mathcal{G}}'^y_{A,j}\mathrm{=}{\mathcal{G}}^y_{A,j}f\left(\mathrm{-}\tau \left(\mathit{\Delta}B_F\left(E_{{\mathrm{L}}_l}\left(j,y\right)\right)\right)\right),                                                                     
\end{equation} 
where 
\begin{equation} \label{6)} 
\begin{split}
\mathit{\Delta}B_F\left(E_{{\mathrm{L}}_l}\left(j,y\right)\right)\\\mathrm{=}&\left|\frac{\sum^n_{h\mathrm{=1}}{B_F\left(E_{{\mathrm{L}}_l}\left(y,h\right)\right)}}{n}\mathrm{-}B_F\left(E_{{\mathrm{L}}_l}\left(j,y\right)\right)\right|,                                                     
\end{split}
\end{equation} 
where $B_F\left(E_{{\mathrm{L}}_l}\left(y,i\right)\right)$ is the entanglement throughput of link $E_{{\mathrm{L}}_l}\left(y,i\right)$ between nodes $y$ and $i$. 

 By some fundamental theory on swarm intelligence [\cref{ref12}-\cref{ref19}], we set the exponential distribution function for $f\left(x\right)$, as 
\begin{equation} \label{7)} 
f\left(x\right)\mathrm{=}e^x,                                                                                          
\end{equation} 
from which \eqref{ZEqnNum489942} is as
\begin{equation} \label{ZEqnNum270243} 
{\mathcal{G}}'^y_{A,x}\mathrm{=}{\mathcal{G}}^y_{A,x}e^{\mathrm{-}\tau \left(\mathit{\Delta}B_F\left(E_{{\mathrm{L}}_l}\left(x,y\right)\right)\right)}\mathrm{+}{\lambda }'_{E_{{\mathrm{L}}_l}\left(x,y\right)},                                                                    
\end{equation} 
while \eqref{ZEqnNum529081} can be rewritten as
\begin{equation} \label{9)} 
{\mathcal{G}}'^y_{A,j}\mathrm{=}{\mathcal{G}}^y_{A,j}e^{\mathrm{-}\tau \left(\mathit{\Delta}B_F\left(E_{{\mathrm{L}}_l}\left(j,y\right)\right)\right)}.                                                                              
\end{equation} 
 
\subsection{Stochastic Model of Entanglement Utility }
Let focus on the ${\mathcal{G}}^y_{A,x}$ evolution (see \eqref{ZEqnNum489942}) at a given ${\lambda }_{E_{{\mathrm{L}}_l}\left(x,y\right)}$ entanglement utility of link $E_{{\mathrm{L}}_l}\left(x,y\right)$ between a current node $y$, and a previous node $x$. Since the entanglement utility of a given link $E_{{\mathrm{L}}_l}\left(x,y\right)$ evolves in time (see \eqref{ZEqnNum640853}) for $E_{{\mathrm{L}}_l}\left(x,y\right)$, the ${\lambda }_{E_{{\mathrm{L}}_l}\left(x,y\right)}$ entanglement utility can be modeled as a non-negative, non-stationary random [\cref{ref15}-\cref{ref16}] process $X^y_{E_{{\mathrm{L}}_l}\left(x,y\right)}\left(t\right)$, with mean ${\mu }^y_{E_{{\mathrm{L}}_l}\left(x,y\right)}\left(t\right)$. As follows, ${\lambda }_{E_{{\mathrm{L}}_l}\left(x,y\right)}$ provides a sample of process $X^y_{E_{{\mathrm{L}}_l}\left(x,y\right)}\left(t\right)$.

 Let $E\left[X^y_{E_{{\mathrm{L}}_l}\left(x,y\right)}\left(t\right)\right]$ be the estimate of $X^y_{E_{{\mathrm{L}}_l}\left(x,y\right)}\left(t\right)$, defined as
\begin{equation} \label{10)} 
E\left[X^y_{E_{{\mathrm{L}}_l}\left(x,y\right)}\left(t\right)\right]\mathrm{=}X^y_{E_{{\mathrm{L}}_l}\left(x,y\right)}\left(t\right)\mathrm{*}{\mathrm{\Omega }}_{{\mathcal{G}}^y_{A,x}}\left(t\right),                                                           
\end{equation} 
where $\mathrm{*}$ is the convolution operator, while function ${\mathrm{\Omega }}_{{\mathcal{G}}^y_{A,x}}\left(t\right)$ is defined as
\begin{equation} \label{ZEqnNum268197} 
{\mathrm{\Omega }}_{{\mathcal{G}}^y_{A,x}}\left(t\right)\mathrm{=}e^{\mathrm{-}\tau \left(\mathit{\Delta}B_F\left(E_{{\mathrm{L}}_l}\left(x,y\right)\right)\right)}U\left(t\right),                                                                      
\end{equation} 
where $U\left(t\right)$is the unit step function. 

Assuming that the individual samples of $X^y_{E_{{\mathrm{L}}_l}\left(x,y\right)}\left(t\right)$ within a time period $\mathrm{\Delta }T$ are determined, a ${\mathrm{\bot }}_{{\mathcal{G}}^y_{A,x}}\left(\mathrm{\Delta }T\right)$  correlation function can be defined as
\begin{equation} \label{ZEqnNum949043} 
{\mathrm{\bot }}_{{\mathcal{G}}^y_{A,x}}\left(\mathrm{\Delta }T\right)\mathrm{=}e^{\mathrm{-}\tau \left|\mathrm{\Delta }T\right|}.                                                                            
\end{equation} 

\subsection{Link Selection Probability}
Using the entanglement gradient ${\mathcal{G}}'^y_{z,B}$ in a current node $y$ with neighbor node $z$, the $\mathrm{P}{\mathrm{r}}^y_{E_{{\mathrm{L}}_l}\left(y,z\right)}$ probability that from node $y$ the entangled link $E_{{\mathrm{L}}_l}\left(y,z\right)$ is selected to reach destination $B$ is defined as 
\begin{equation} \label{ZEqnNum536222} 
\begin{split}
\mathrm{P}{\mathrm{r}}^y_{E_{{\mathrm{L}}_l}\left(y,z\right)}\\\mathrm{=}&\frac{{\left({\mathcal{G}}'^y_{z,B}\mathrm{+}\mathrm{\partial }\right)}^{\chi }}{\sum_k{{\left({\mathcal{G}}'^y_{k,B}\mathrm{+}\mathrm{\partial }\right)}^{\chi }}}\\\mathrm{=}&\frac{{\left(\left({\mathcal{G}}^y_{z,B}e^{\mathrm{-}\tau \left(\mathit{\Delta}B_F\left(E_{{\mathrm{L}}_l}\left(y,z\right)\right)\right)}\mathrm{+}{\lambda }'_{E_{{\mathrm{L}}_l}\left(y,z\right)}\right)\mathrm{+}\mathrm{\partial }\right)}^{\chi }}{\sum_k{{\left(\left({\mathcal{G}}^y_{k,B}e^{\mathrm{-}\tau \left(\mathit{\Delta}B_F\left(E_{{\mathrm{L}}_l}\left(y,k\right)\right)\right)}\right)\mathrm{+}\mathrm{\partial }\right)}^{\chi }}},                                
\end{split}
\end{equation} 
where $k\mathrm{\in }V\mathrm{-}y$, $V$ is the set of nodes of the entangled quantum network $N$, $\mathrm{\partial }\mathrm{\ge }\mathrm{0}$ is a threshold parameter, while $\chi \mathrm{\ge }\mathrm{0}$ is a tuning parameter. A source-dependent link selection model is discussed in \sref{secA1}.

\section{Path Entanglement-Gradient}
\label{sec3}
 The relevance of a particular path of the network is characterized by the path entanglement gradient coefficient. 

 In this section, we extend the entanglement gradient to entangled paths, which refers to the paths between source and target nodes in the quantum network that are formed by a chain of entangled links between quantum repeaters (i.e., paths of entangled links).  

\begin{figure*}[h!]
\vspace{-0.6cm}
 \begin{center}
 	 \includegraphics[angle = 0,width=1\linewidth]{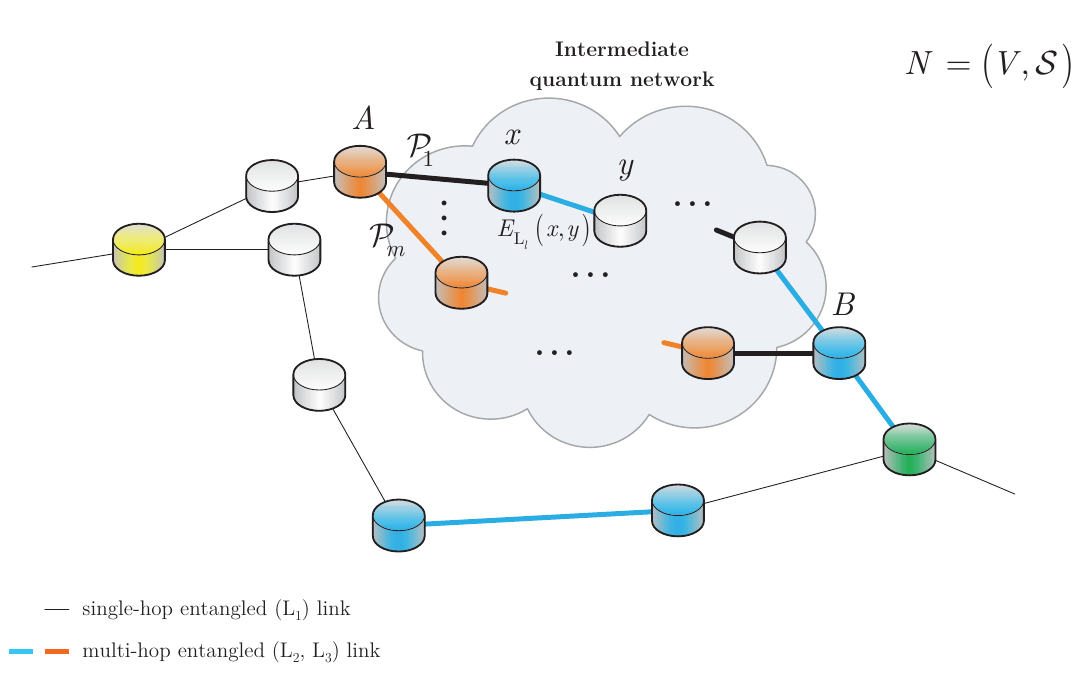}

\caption{A quantum network with source node $A$ and destination node $B$, and $m$ entangled paths ${\mathcal{P}}_{\mathrm{1}}\mathrm{,\dots ,}{\mathcal{P}}_m$ between them. Each path is formulated by a chain of entangled links between quantum repeater nodes. The actual network topology between $A$ and $B$ is unknown (depicted by the cloud) and paths ${\mathcal{P}}_{\mathrm{1}}\mathrm{,\dots ,}{\mathcal{P}}_m$ abstract all entangled links and noise between $A$ and $B$. A section of path ${\mathcal{P}}_{\mathrm{1}}$ is illustrated by an ${\mathrm{L}}_l$-level entangled link $E_{{\mathrm{L}}_l}\left(x,y\right)$ between nodes $\left(x,y\right)$ of the particular path.}
 \label{fig1}
\end{center}
\end{figure*}

 The network model used for the entanglement-gradient routing scheme is illustrated in \fref{fig1}. There are $m$ entangled paths, ${\mathcal{P}}_{\mathrm{1}}\mathrm{,\dots ,}{\mathcal{P}}_m$  between a source node $A$ and destination node $B$. Each entangled path ${\mathcal{P}}_i$, $i\mathrm{=1,\dots ,}m$, is formulated by a chain of entangled links between quantum repeaters.

\subsection{Path Metrics }

 In this section, we focus on the entanglement gradients of the $m$ entangled paths ${\mathcal{P}}_{\mathrm{1}}\mathrm{,\dots ,}{\mathcal{P}}_m$ between a source node $A$ and target node $B$. 

 Let ${\mathcal{G}}^A_{{\mathcal{P}}_i}$ refer to the initial path entanglement gradient of a given entangled path ${\mathcal{P}}_i$, $i\mathrm{=1,\dots ,}m$ at source node $A$. Let ${\mathcal{G}}^B_{{\mathcal{P}}_i}$ be the initial path entanglement gradient of ${\mathcal{P}}_i$ at destination node $B$ [\cref{ref15}-\cref{ref18}]. Let ${\kappa }_A$ be the mean number of $d$-dimensional entangled states arriving at $A$ and ${\kappa }_B$ be the mean number arriving at $B$; therefore, the total observation rate is
\begin{equation} \label{14)} 
{\kappa }_{AB}\mathrm{=}{\kappa }_A\mathrm{+}{\kappa }_B.                                                                                        
\end{equation} 
Note that, assuming a symmetrical arrival of the entangled states, ${\kappa }_A\mathrm{=}{\kappa }_B\\\mathrm{=}{{\kappa }_{AB}}/{\mathrm{2}}$.

 The derivation of updated ${\mathcal{G}}'^A_{{\mathcal{P}}_i}$ at the source node $A$ for a given path ${\mathcal{P}}_i$ is as follows. Let ${\mathcal{G}}^A_{{\mathcal{P}}_i}$ be the initial gradient in $A$ and let ${\mathcal{P}}_i$, characterized by a $X^A_{{\mathcal{P}}_i}\left(t\right)$, be the non-stationary random process with mean ${\mu }^A_{{\mathcal{P}}_i}$ (average value of received entanglement gradient).

 First, ${\mathcal{G}}'^A_{{\mathcal{P}}_i}$ is decomposed to
\begin{equation} \label{ZEqnNum919296} 
{\mathcal{G}}'^A_{{\mathcal{P}}_i}\mathrm{=}{\mathcal{G}}^{A,\left({\kappa }_A,{\tau }_A\right)}_{{\mathcal{P}}_i}\mathrm{+}{\mathcal{G}}^{A,\left({\mathcal{P}}_i\right)}_{{\mathcal{P}}_i}\mathrm{+}{\mathcal{G}}^{A,\left({\mathcal{P}}_j\right)}_{{\mathcal{P}}_i},                                                                
\end{equation} 
where the first term, ${\mathcal{G}}^{A,\left({\kappa }_A,{\tau }_A\right)}_{{\mathcal{P}}_i}$, is the entanglement gradient update in $A$, evaluated as
\begin{equation} \label{ZEqnNum311795} 
{\mathcal{G}}^{A,\left({\kappa{e}}_A,{\tau }_A\right)}_{{\mathcal{P}}_i}\mathrm{=}\frac{{\kappa }_A}{{\kappa }_{AB}}\left(\frac{{\kappa }_{AB}}{{\kappa }_{AB}\mathrm{+}{\tau }_A}\right){\mathcal{G}}^A_{{\mathcal{P}}_i},                                                                       
\end{equation} 
where ${\tau }_A$ is the decay rate of $A$.   

 The second term ${\mathcal{G}}^{A,\left({\mathcal{P}}_i\right)}_{{\mathcal{P}}_i}$ models the entanglement gradient update for the given path ${\mathcal{P}}_i$ as
\begin{equation} \label{ZEqnNum246466} 
{\mathcal{G}}^{A,\left({\mathcal{P}}_i\right)}_{{\mathcal{P}}_i}\mathrm{=P}{\mathrm{r}}^B_{{\mathcal{P}}_i}\left(\frac{{\kappa }_B}{{\kappa }_{AB}}\left(\left(\frac{{\kappa }_{AB}}{{\kappa }_{AB}\mathrm{+}{\tau }_A}\right){\mathcal{G}}^A_{{\mathcal{P}}_i}\mathrm{+}\left({\mu }^A_{{\mathcal{P}}_i}\right)\right)\right),                                                            
\end{equation} 
where ${\mu }^A_{{\mathcal{P}}_i}$ is the average value of received entanglement gradient from path ${\mathcal{P}}_i$ at node $A$, while $\mathrm{P}{\mathrm{r}}^B_{{\mathcal{P}}_i}$ is the probability that path ${\mathcal{P}}_i$ will be used by $B$, $\mathrm{P}{\mathrm{r}}^B_{{\mathcal{P}}_i}\mathrm{=}{{\left({\mathcal{G}}^B_{{\mathcal{P}}_i}\mathrm{+}\mathrm{\partial }\right)}^{\chi }}/{{\sum_m{\left({\mathcal{G}}^B_{{\mathcal{P}}_i}\mathrm{+}\mathrm{\partial }\right)}}^{\chi }}$.                                             

 The third term, ${\mathcal{G}}^{A,\left({\mathcal{P}}_j\right)}_{{\mathcal{P}}_i}$ models the entanglement gradient update for a different path ${\mathcal{P}}_j$, $j\mathrm{\ne }i$ as
\begin{equation} \label{ZEqnNum241844} 
{\mathcal{G}}^{A,\left({\mathcal{P}}_j\right)}_{{\mathcal{P}}_i}\mathrm{=P}{\mathrm{r}}^B_{{\mathcal{P}}_j}\left(\frac{{\kappa }_B}{{\kappa }_{AB}}\left(\left(\frac{{\kappa }_{AB}}{{\kappa }_{AB}\mathrm{+}{\tau }_A}\right){\mathcal{G}}^A_{{\mathcal{P}}_i}\right)\right),                                                                 
\end{equation} 
where $\mathrm{P}{\mathrm{r}}^B_{{\mathcal{P}}_j}$ is the probability that ${\mathcal{P}}_j$ will be used by $B$, $\mathrm{P}{\mathrm{r}}^B_{{\mathcal{P}}_j}\mathrm{=}{{\left({\mathcal{G}}^B_{{\mathcal{P}}_j}\mathrm{+}\mathrm{\partial }\right)}^{\chi }}/\\{{\sum_m{\left({\mathcal{G}}^B_{{\mathcal{P}}_i}\mathrm{+}\mathrm{\partial }\right)}}^{\chi }}$. 

 From \eqref{ZEqnNum311795}, \eqref{ZEqnNum246466}, and \eqref{ZEqnNum241844}, ${\mathcal{G}}'^A_{{\mathcal{P}}_i}$ in \eqref{ZEqnNum919296} can be rewritten for a particular path ${\mathcal{P}}_i$ as
\begin{equation} \label{ZEqnNum937895} 
{\mathcal{G}}'^A_{{\mathcal{P}}_i}\mathrm{=}\left(\frac{{\kappa }_{AB}}{{\kappa }_{AB}\mathrm{+}{\tau }_A}\right){\mathcal{G}}^A_{{\mathcal{P}}_i}\mathrm{+}\left(\frac{{\kappa }_B}{{\kappa }_{AB}}\right)\mathrm{P}{\mathrm{r}}^B_{{\mathcal{P}}_i}\left({\mu }^A_{{\mathcal{P}}_i}\right). 
\end{equation} 
Following the same steps for path ${\mathcal{P}}_j$, $j\mathrm{\ne }i$, ${\mathcal{G}}'^A_{{\mathcal{P}}_j}$ is evaluated at node $A$, for all instances of $j$, as
\begin{equation} \label{ZEqnNum364941} 
{\mathcal{G}}'^A_{{\mathcal{P}}_j}\mathrm{=}\left(\frac{{\kappa }_{AB}}{{\kappa }_{AB}\mathrm{+}{\tau }_A}\right){\mathcal{G}}^A_{{\mathcal{P}}_j}\mathrm{+}\left(\frac{{\kappa }_B}{{\kappa }_{AB}}\right)\mathrm{P}{\mathrm{r}}^B_{{\mathcal{P}}_j}\left({\mu }^A_{{\mathcal{P}}_j}\right). 
\end{equation} 
At target node $B$, the corresponding formula for path ${\mathcal{P}}_i$, ${\mathcal{G}}'^B_{{\mathcal{P}}_i}$ is therefore yielded as
\begin{equation} \label{21)} 
{\mathcal{G}}'^B_{{\mathcal{P}}_i}\mathrm{=}\left(\frac{{\kappa }_{AB}}{{\kappa }_{AB}\mathrm{+}{\tau }_B}\right){\mathcal{G}}^B_{{\mathcal{P}}_i}\mathrm{+}\left(\frac{{\kappa }_A}{{\kappa }_{AB}}\right)\mathrm{P}{\mathrm{r}}^A_{{\mathcal{P}}_i}\left({\mu }^B_{{\mathcal{P}}_i}\right), 
\end{equation} 
where ${\mu }^B_{{\mathcal{P}}_i}$ is the average value of received entanglement gradient from path ${\mathcal{P}}_i$ at node $B$, $\mathrm{P}{\mathrm{r}}^A_{{\mathcal{P}}_i}$ is the probability that path ${\mathcal{P}}_i$ will be used by $A$, and $\mathrm{P}{\mathrm{r}}^A_{{\mathcal{P}}_i}\mathrm{=}{{\left({\mathcal{G}}^A_{{\mathcal{P}}_i}\mathrm{+}\mathrm{\partial }\right)}^{\chi }}/{{\sum_m{\left({\mathcal{G}}^A_{{\mathcal{P}}_i}\mathrm{+}\mathrm{\partial }\right)}}^{\chi }}$. 

 The formula of ${\mathcal{G}}'^B_{{\mathcal{P}}_j}$ for path ${\mathcal{P}}_j$, $j\mathrm{\ne }i$ at target node $B$ is therefore
\begin{equation} \label{22)} 
{\mathcal{G}}'^B_{{\mathcal{P}}_j}\mathrm{=}\left(\frac{{\kappa }_{AB}}{{\kappa }_{AB}\mathrm{+}{\tau }_B}\right){\mathcal{G}}^B_{{\mathcal{P}}_j}\mathrm{+}\left(\frac{{\kappa }_A}{{\kappa }_{AB}}\right)\mathrm{P}{\mathrm{r}}^A_{{\mathcal{P}}_j}\left({\mu }^B_{{\mathcal{P}}_j}\right), 
\end{equation} 
where $\mathrm{P}{\mathrm{r}}^A_{{\mathcal{P}}_j}$ is $\mathrm{P}{\mathrm{r}}^A_{{\mathcal{P}}_j}\mathrm{=}{{\left({\mathcal{G}}^A_{{\mathcal{P}}_j}\mathrm{+}\mathrm{\partial }\right)}^{\chi }}/{{\sum_m{\left({\mathcal{G}}^A_{{\mathcal{P}}_i}\mathrm{+}\mathrm{\partial }\right)}}^{\chi }}$.

 For the ${\mathcal{P}}^{\mathrm{*}}$ optimal shortest path, the entanglement gradient is maximal, thus ${\mathcal{G}}'^A_{{\mathcal{P}}^{\mathrm{*}}}$ is determined as
\begin{equation} \label{ZEqnNum282699} 
 \begin{array}{l}
\begin{split}
{\mathcal{G}}'^A_{{\mathcal{P}}^{\mathrm{*}}}&\mathrm{=}\mathop{\mathrm{max}}_{\mathrm{\forall }i}{\mathcal{G}}'^A_{{\mathcal{P}}_i} \\ 
&\mathrm{=}\left(\frac{{\kappa }_{AB}}{{\kappa }_{AB}\mathrm{+}{\tau }_A}\right){\mathcal{G}}^A_{{\mathcal{P}}^{\mathrm{*}}}\mathrm{+}\left(\frac{{\kappa }_B}{{\kappa }_{AB}}\right)\mathrm{P}{\mathrm{r}}^B_{{\mathcal{P}}^{\mathrm{*}}}\left({\mu }^A_{{\mathcal{P}}^{\mathrm{*}}}\right). 
\end{split}
\end{array}
\end{equation} 
 
\subsubsection{Mean of path entanglement gradient}

 After some calculations, the mean entanglement gradient $\mathbb{E}\left({\mathcal{G}}'^A_{{\mathcal{P}}_i}\right)$ of a particular path ${\mathcal{P}}_i$ at $A$ is obtainable if the path ${\mathcal{P}}_i$ is selected with unit probability in $A$, $\mathrm{P}{\mathrm{r}}^A_{{\mathcal{P}}_i}\mathrm{=1}$, yielding $\mathbb{E}\left({\mathcal{G}}'^A_{{\mathcal{P}}_i}\right)$ as
\begin{equation} \label{ZEqnNum406725} 
\mathbb{E}\left({\mathcal{G}}'^A_{{\mathcal{P}}_i}\right)\mathrm{=}\frac{\left({\kappa }_{AB}\mathrm{+}{\tau }_A\right){\kappa }_B}{{\kappa }_{AB}{\tau }_A}{\mu }^A_{{\mathcal{P}}_i}.                                                                           
\end{equation} 
By similar assumptions, the $\mathbb{E}\left({\mathcal{G}}'^B_{{\mathcal{P}}_i}\right)$ mean entanglement gradient of a particular path ${\mathcal{P}}_i$ at $B$, obtainable at $\mathrm{P}{\mathrm{r}}^B_{{\mathcal{P}}_i}\mathrm{=1}$, is
\begin{equation} \label{ZEqnNum485557} 
\mathbb{E}\left({\mathcal{G}}'^B_{{\mathcal{P}}_i}\right)\mathrm{=}\frac{\left({\kappa }_{AB}\mathrm{+}{\tau }_B\right){\kappa }_A}{{\kappa }_{AB}{\tau }_B}{\mu }^B_{{\mathcal{P}}_i}.                                                                         
\end{equation}

\subsection{Decay Rate of Mean Path Entanglement Gradient}

 A crucial parameter for the optimization of the entanglement-gradient routing is the ${\tau }_{{\mathcal{G}}'^n_{{\mathcal{P}}_i}}$ decay rate [\cref{ref15}-\cref{ref18}] of mean path entanglement gradient $\mathbb{E}\left({\mathcal{G}}'^n_{{\mathcal{P}}_i}\right)$. 

 Without loss of generality, at a given expected amount of entanglement gradient $\mathbb{E}\left({\mathcal{G}}'^n_{{\mathcal{P}}_i}\right)$ at node $n$ for path ${\mathcal{P}}_i$, the threshold ${\mathrm{\partial }}_{\mathbb{E}\left({\mathcal{G}}'^n_{{\mathcal{P}}_i}\right)}$ can be rewritten as:
\begin{equation} \label{ZEqnNum478550} 
{\mathrm{\partial }}_{\mathbb{E}\left({\mathcal{G}}'^n_{{\mathcal{P}}_i}\right)}\mathrm{=}\mathbb{E}\left({\mathcal{G}}'^n_{{\mathcal{P}}_i}\right)e^{\mathrm{-}\varphi \left({\mathcal{P}}_i\right){\tau }_{\mathbb{E}\left({\mathcal{G}}'^n_{{\mathcal{P}}_i}\right)}}, 
\end{equation} 
where $\varphi \left({\mathcal{P}}_i\right)$ characterizes the deviation of a current $B_F\left({\mathcal{P}}_i\right)$ entanglement throughput (measured in $d$-dimensional entangled states of a particular fidelity $F$ per sec) of path ${\mathcal{P}}_i$ from an expected ${\tilde{B}}_F\left({\mathcal{P}}_i\right)$ entanglement throughput of path ${\mathcal{P}}_i$ as
\begin{equation} \label{ZEqnNum645414} 
\varphi \left({\mathcal{P}}_i\right)\mathrm{=}\left|{\tilde{B}}_F\left({\mathcal{P}}_i\right)\mathrm{-}B_F\left({\mathcal{P}}_i\right)\right|.                                                                  
\end{equation} 
and therefore ${\tau }_{\mathbb{E}\left({\mathcal{G}}'^n_{{\mathcal{P}}_i}\right)}$ is
\begin{equation} \label{ZEqnNum763249} 
\begin{split}
{\tau }_{\mathbb{E}\left({\mathcal{G}}'^n_{{\mathcal{P}}_i}\right)}\\&\mathrm{=-ln}\frac{{\mathrm{\partial }}_{\mathbb{E}\left({\mathcal{G}}'^n_{{\mathcal{P}}_i}\right)}}{\mathbb{E}\left({\mathcal{G}}'^n_{{\mathcal{P}}_i}\right)\varphi \left({\mathcal{P}}_i\right)}\\&\mathrm{=-ln}\frac{{\mathrm{\partial }}_{\mathbb{E}\left({\mathcal{G}}'^n_{{\mathcal{P}}_i}\right)}}{\mathbb{E}\left({\mathcal{G}}'^n_{{\mathcal{P}}_i}\right)\left|{\tilde{B}}_F\left({\mathcal{P}}_i\right)\mathrm{-}B_F\left({\mathcal{P}}_i\right)\right|}.                                             
\end{split}
\end{equation} 
 
\subsubsection{Optimal estimator}

 The ${\widetilde{\tau }}_{\mathbb{E}\left({\mathcal{G}}'^n_{{\mathcal{P}}_i}\right)}$ optimal estimator of ${\tau }_{\mathbb{E}\left({\mathcal{G}}'^n_{{\mathcal{P}}_i}\right)}$ is derived as follows. Using \eqref{ZEqnNum949043} with \eqref{ZEqnNum645414} allows us to evaluate a variable $Y$ as
\begin{equation} \label{29)} 
Y\mathrm{=\ \ }{\mathrm{\bot }}_{\mathbb{E}\left({\mathcal{G}}'^n_{{\mathcal{P}}_i}\right)}\left(\varphi \left({\mathcal{P}}_i\right)\right)\mathrm{=}e^{\mathrm{-}{\widetilde{\tau }}_{\mathbb{E}\left({\mathcal{G}}'^n_{{\mathcal{P}}_i}\right)}\varphi \left({\mathcal{P}}_i\right)},                                                           
\end{equation} 
from which the ${\widetilde{\tau }}_{\mathbb{E}\left({\mathcal{G}}'^n_{{\mathcal{P}}_i}\right)}$ optimal estimate of the ${\tau }_{\mathbb{E}\left({\mathcal{G}}'^n_{{\mathcal{P}}_i}\right)}$ of $\mathbb{E}\left({\mathcal{G}}'^n_{{\mathcal{P}}_i}\right)$ is yielded as [\cref{ref15}-\cref{ref18}]
\begin{equation} \label{ZEqnNum508033} 
\begin{split}
{\widetilde{\tau }}_{\mathbb{E}\left({\mathcal{G}}'^n_{{\mathcal{P}}_i}\right)}\\&\mathrm{=-}\frac{\mathrm{ln}Y}{\varphi \left({\mathcal{P}}_i\right)}\\&\mathrm{=-}\frac{\mathrm{ln}\left(e^{\mathrm{-}{\widetilde{\tau }}_{\mathbb{E}\left({\mathcal{G}}'^n_{{\mathcal{P}}_i}\right)}\varphi \left({\mathcal{P}}_i\right)}\right)}{\varphi \left({\mathcal{P}}_i\right)}.                                                          
\end{split}
\end{equation} 
At a given optimal decay rate ${\widetilde{\tau }}_{\mathbb{E}\left({\mathcal{G}}'^n_{{\mathcal{P}}_i}\right)}$ \eqref{ZEqnNum508033}, using \eqref{ZEqnNum406725} in \eqref{ZEqnNum478550} results in ${\mathrm{\partial }}_{\mathbb{E}\left({\mathcal{G}}'^n_{{\mathcal{P}}_i}\right)}$ as 
\begin{equation} \label{31)} 
\begin{split}
{\mathrm{\partial }}_{\mathbb{E}\left({\mathcal{G}}'^n_{{\mathcal{P}}_i}\right)}\\&\mathrm{=}\frac{\left({\kappa }_{AB}\mathrm{+}{\widetilde{\tau }}_{\mathbb{E}\left({\mathcal{G}}'^n_{{\mathcal{P}}_i}\right)}\right)}{\mathrm{2}{\widetilde{\tau }}_{\mathbb{E}\left({\mathcal{G}}'^n_{{\mathcal{P}}_i}\right)}}{\mu }^n_{{\mathcal{P}}_i}e^{\mathrm{-}{\widetilde{\tau }}_{\mathbb{E}\left({\mathcal{G}}'^n_{{\mathcal{P}}_i}\right)}\varphi \left({\mathcal{P}}_i\right)}.                                                            
\end{split}
\end{equation} 
 
\subsection{Path Selection}

A brief description of the method to determine the entanglement gradient of the paths for characterization of an optimal path ${\mathcal{P}}^{\mathrm{*}}$ is summarized in Method 1. 

\begin{method}
  \DontPrintSemicolon
\caption{\textit{Path entanglement gradient}}
\textbf{Step 1.} Let $n\mathrm{-}\mathrm{1}$ and $n$ be a pair of neighbor quantum repeaters of a path between source node $A$ and target node $B$. 
Let $n$ be the current node, $n\mathrm{-}\mathrm{1}$ be the previous node, and $n\mathrm{+1}$ be a next node. 

\textbf{Step 2.} Apply \eqref{ZEqnNum640853} to increase the entanglement utility of the entangled link between  $n\mathrm{-}\mathrm{1}$ and $n$. For node $n\mathrm{-}\mathrm{1}$, increase entanglement gradient via \eqref{ZEqnNum489942}. For all other neighboring nodes, decrease entanglement gradient via \eqref{ZEqnNum529081}.

\textbf{Step 3.} From the updated entanglement gradients, compute $\mathrm{P}{\mathrm{r}}^n_{E_{{\mathrm{L}}_l}\left(n,n\mathrm{+1}\right)}$ of entangled link $E_{{\mathrm{L}}_l}\left(n,n\mathrm{+1}\right)$ via \eqref{ZEqnNum536222}.

\textbf{Step 4.} Apply steps 1--3 for all nodes and paths, ${\mathcal{P}}_{\mathrm{1}}\mathrm{,\dots ,}{\mathcal{P}}_m$. Determine optimal $\widetilde{\tau }$ via \eqref{ZEqnNum508033} to set $\tau $. 

\textbf{Step 5.} Using \eqref{ZEqnNum282699}, output optimal path ${\mathcal{P}}^{\mathrm{*}}$ for which the entanglement gradient is maximal is ${\mathcal{G}}'^A_{{\mathcal{P}}^{\mathrm{*}}}\mathrm{=}\mathop{\mathrm{max}}_{\mathrm{\forall }i}{\mathcal{G}}'^A_{{\mathcal{P}}_i}$. 
\end{method}
 
\section{Entanglement-Gradient Routing }
\label{sec4}
In this section, we define a decentralized routing scheme that merges the results of the previous sections on the quantities of entanglement gradient. The routing is executed through parallel threads that simultaneously explore the quantum network. A given thread operates in a localized manner.
 
\subsection{Link Selection}

 For a given path ${\mathcal{P}}_i$ between a source node $s$ and current node $n$, a quantity ${\mathrm{\Phi }}^{s,n}_{{\mathcal{P}}_i}$ is defined as 
\begin{equation} \label{ZEqnNum650027} 
{\mathrm{\Phi }}^{s,n}_{{\mathcal{P}}_i}\mathrm{=}\sum^n_{x\mathrm{=}s}{\alpha {\sigma }^x_{{\mathcal{P}}_i}},                                                                                
\end{equation} 
where
\begin{equation} \label{ZEqnNum212998} 
{\sigma }^x_{{\mathcal{P}}_i}\mathrm{=log}\left(\frac{{\mathcal{G}}'^{x\mathrm{+1}\mathrm{\in }{\mathcal{P}}_i}_{{\mathcal{P}}_i}}{{\mathcal{G}}'^{x\mathrm{\in }{\mathcal{P}}_i}_{{\mathcal{P}}_i}}\right),                                                                             
\end{equation} 
where ${\mathcal{G}}'^{x\mathrm{\in }{\mathcal{P}}_i}_{{\mathcal{P}}_k}$ is the entanglement gradient of node $x\mathrm{\in }{\mathcal{P}}_i$, ${\mathcal{G}}'^{x\mathrm{+1}\mathrm{\in }{\mathcal{P}}_i}_{{\mathcal{P}}_k}$ is the entanglement gradient at node $x\mathrm{+1}\mathrm{\in }{\mathcal{P}}_i$, and $\alpha $ is
\begin{equation} \label{ZEqnNum549282} 
\alpha \mathrm{=}\left\{ \begin{array}{l}
\mathrm{1,}\textnormal{ if }\left|{\sigma }^x_{{\mathcal{P}}_i}\right|\mathrm{>}\vartheta  \\ 
0,\textnormal{ if }\left|{\sigma }^x_{{\mathcal{P}}_i}\right|\mathrm{\le }\vartheta  \end{array}
\right.,                                                                         
\end{equation} 
where $\vartheta $ is a threshold [\cref{ref17}]. 

\begin{figure*}[h!]
\vspace{-0.6cm}
 \begin{center}
 	 \includegraphics[angle = 0,width=1\linewidth]{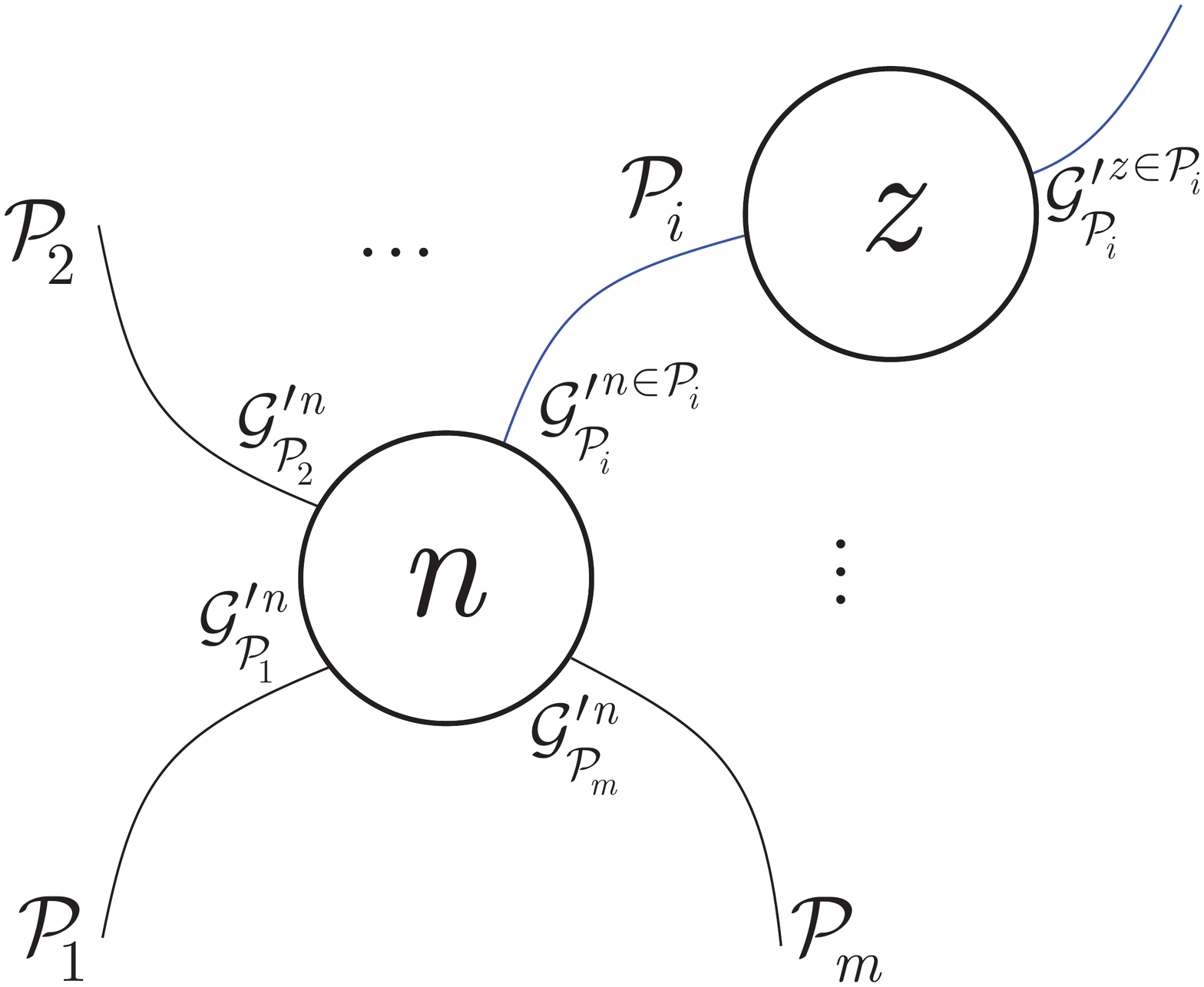}

\caption{A model of $m$ paths ${\mathcal{P}}_{\mathrm{1}}\mathrm{,\dots ,}{\mathcal{P}}_m$ between a source node $s$ and a current node $n$. The paths have entanglement gradients ${\mathcal{G}}'^n_{{\mathcal{P}}_i}$, $i\mathrm{=1,}\mathrm{\dots ,}m$. Current node $n$ and next node $z$ are elements of a current path ${\mathcal{P}}_i$ (blue), with ${\mathcal{G}}'^{n\mathrm{\in }{\mathcal{P}}_i}_{{\mathcal{P}}_i}$ in $n$ and ${\mathcal{G}}'^{z\mathrm{\in }{\mathcal{P}}_i}_{{\mathcal{P}}_i}$ in $z$.}
\label{fig2}
\end{center}
\end{figure*}

 Using \eqref{ZEqnNum650027}, a mean ${\mu }^n\left({\mathrm{\Phi }}^{s,n}_{\mathcal{P}}\right)$ for the $m$ paths ${\mathcal{P}}_{\mathrm{1}}\mathrm{,\dots ,}{\mathcal{P}}_m$ between a source node $s$ and a current node $n$ is
\begin{equation} \label{ZEqnNum751770} 
{\mu }^n\left({\mathrm{\Phi }}^{s,n}_{\mathcal{P}}\right)\mathrm{=}\frac{\sum^m_{i\mathrm{=1}}{{\mathrm{\Phi }}^{s,n}_{{\mathcal{P}}_i}}}{m}.                                                                       
\end{equation} 
A model of a node $n$ with next node $z$ and $m$ paths ${\mathcal{P}}_{\mathrm{1}}\mathrm{,\dots ,}{\mathcal{P}}_m$ between a source node $s$ is depicted in \fref{fig2}. The entanglement gradients are ${\mathcal{G}}'^n_{{\mathcal{P}}_i}$, $i\mathrm{=1,\dots ,}m$. Nodes $n$ and $z$ are elements of a current path ${\mathcal{P}}_i$, with corresponding entanglement gradients ${\mathcal{G}}'^{n\mathrm{\in }{\mathcal{P}}_i}_{{\mathcal{P}}_i}$ and ${\mathcal{G}}'^{z\mathrm{\in }{\mathcal{P}}_i}_{{\mathcal{P}}_i}$. From the path gradients, the quantities of ${\sigma }^x_{{\mathcal{P}}_i}$ \eqref{ZEqnNum212998} and $\alpha $ \eqref{ZEqnNum549282} are derived to evaluate ${\mathrm{\Phi }}^{s,n}_{{\mathcal{P}}_i}$ in \eqref{ZEqnNum650027}.
 \\

Let $z$ be the next node from actual node $n$ on a current path ${\mathcal{P}}_i$ with entangled connection $E_{{\mathrm{L}}_l}\left(n,z\right)$. Then, for $E_{{\mathrm{L}}_l}\left(n,z\right)$, the distance function $\psi \left(n,z\right)$ between $n$ and $z$ is defined as
\begin{equation} \label{ZEqnNum867276} 
 \begin{array}{l}
\psi \left(n,z\right)\\\mathrm{=}\left(\frac{\mathrm{1}}{\mathrm{P}{\mathrm{r}}^n_{E_{{\mathrm{L}}_l}\left(n,z\right)}}\right)\left|\mathbb{E}\left({\mathcal{G}}'^n_{{\mathcal{P}}_i}\right)\mathrm{-}\mathbb{E}\left({\mathcal{G}}'^z_{{\mathcal{P}}_i}\right)\right| \\ 
\\\mathrm{=}\left(\frac{\sum_k{{\left({\mathcal{G}}'^n_{k,B}\mathrm{+}\mathrm{\partial }\right)}^{\chi }}}{{\left({\mathcal{G}}'^n_{z,B}\mathrm{+}\mathrm{\partial }\right)}^{\chi }}\right)\left|\mathbb{E}\left({\mathcal{G}}'^n_{{\mathcal{P}}_i}\right)\mathrm{-}\mathbb{E}\left({\mathcal{G}}'^z_{{\mathcal{P}}_i}\right)\right|, \end{array}
\end{equation} 
where $\mathrm{P}{\mathrm{r}}^n_{E_{{\mathrm{L}}_l}\left(n,z\right)}$ is \eqref{ZEqnNum536222}, while $\mathbb{E}\left({\mathcal{G}}'^n_{{\mathcal{P}}_i}\right)$ and $\mathbb{E}\left({\mathcal{G}}'^z_{{\mathcal{P}}_i}\right)$ are the mean entanglement gradients at nodes $n\mathrm{\in }{\mathcal{P}}_i$ and $z\mathrm{\in }{\mathcal{P}}_i$, evaluated via \eqref{ZEqnNum406725} and \eqref{ZEqnNum485557} as
\begin{equation} \label{37)} 
\mathbb{E}\left({\mathcal{G}}'^n_{{\mathcal{P}}_i}\right)\mathrm{=}\frac{\left({\kappa }_{nz}\mathrm{+}{\tau }_n\right){\kappa }_z}{{\kappa }_{nz}{\tau }_n}{\mu }^n_{{\mathcal{P}}_i},                                                                     
\end{equation} 
where ${\mu }^n_{{\mathcal{P}}_i}$ is the average value of the received entanglement gradient from path ${\mathcal{P}}_i$ at node $n$, and
\begin{equation} \label{38)} 
\mathbb{E}\left({\mathcal{G}}'^z_{{\mathcal{P}}_i}\right)\mathrm{=}\frac{\left({\kappa }_{nz}\mathrm{+}{\tau }_z\right){\kappa }_n}{{\kappa }_{nz}{\tau }_z}{\mu }^z_{{\mathcal{P}}_i},                                                                       
\end{equation} 
where ${\mu }^z_{{\mathcal{P}}_i}$ is the average value of the received entanglement gradient from path ${\mathcal{P}}_i$ at node $z$, respectively.

 The decentralized routing is accomplished via $t$ parallel threads, ${\mathcal{T}}_{\mathrm{1}}\mathrm{,\dots ,}{\mathcal{T}}_t$. For all threads, a threshold ${\ell }_{\mathcal{T}}$ is defined, which determines the maximal number of nodes to be explored. Using the ${\mathcal{G}}'^z_{A,n}$ entanglement link gradient (see \eqref{ZEqnNum270243}), with the entanglement utility ${\lambda }'_{E_{{\mathrm{L}}_l}\left(n,z\right)}$ \eqref{ZEqnNum640853} of link $E_{{\mathrm{L}}_l}\left(n,z\right)$ between nodes $n$ and $z$, an inverse link entanglement gradient ${\theta }^z_{E_{{\mathrm{L}}_l}\left(n,z\right)}$ is defined as
\begin{equation} \label{ZEqnNum356733} 
{\theta }^z_{E_{{\mathrm{L}}_l}\left(n,z\right)}\mathrm{=}\frac{\mathrm{1}}{{\mathcal{G}}'^z_{A,n}}\mathrm{=}\frac{\mathrm{1}}{{\mathcal{G}}^z_{A,n}e^{\mathrm{-}\tau \left(\mathit{\Delta}B_F\left(E_{{\mathrm{L}}_l}\left(n,z\right)\right)\right)}\mathrm{+}{\lambda }'_{E_{{\mathrm{L}}_l}\left(n,z\right)}},                                                
\end{equation} 
where ${\lambda }'_{E_{{\mathrm{L}}_l}\left(n,z\right)}$ is
\begin{equation} \label{40)} 
{\lambda }'_{E_{{\mathrm{L}}_l}\left(n,z\right)}\mathrm{=}\frac{{\lambda }_{E_{{\mathrm{L}}_l}\left(n,z\right)}}{\mathrm{1+}B_F\left(E_{{\mathrm{L}}_l}\left(n,z\right)\right){\lambda }_{E_{{\mathrm{L}}_l}\left(n,z\right)}}. 
\end{equation} 
Then, for a given $i$-th thread ${\mathcal{T}}_i$, the $p_{{\mathcal{T}}_i}\left(n,z\right)$ link selection probability is defined as
\begin{equation} \label{ZEqnNum706918} 
p_{{\mathcal{T}}_i}\left(n,z\right)\mathrm{=}\left\{ \begin{array}{l}
\mathrm{P}{\mathrm{r}}_{{\mathcal{T}}_i}\left(n,z\right),\textnormal{ if }z\mathrm{\notin }{\mathrm{S}}_{{\mathcal{T}}_i} \\ 
\mathrm{0,otherwise,} \end{array}
\right. 
\end{equation} 
where ${\mathrm{S}}_{{\mathcal{T}}_i}$ is a set of nodes already visited by the $i$-th thread ${\mathcal{T}}_i$ [\cref{ref17}], while $\mathrm{P}{\mathrm{r}}_{{\mathcal{T}}_i}\left(n,z\right)$ is
\begin{equation} \label{ZEqnNum252838} 
\mathrm{P}{\mathrm{r}}_{{\mathcal{T}}_i}\left(n,z\right)\mathrm{=}\frac{{\left({\theta }^z_{E_{{\mathrm{L}}_l}\left(n,z\right)}\right)}^{C_{\mathrm{1}}}{\left(\psi \left(n,z\right)\right)}^{C_{\mathrm{2}}}}{\sum_{k\mathrm{\notin }{\mathrm{S}}_{{\mathcal{T}}_i}}{{\left({\theta }^k_{E_{{\mathrm{L}}_l}\left(n,k\right)}\right)}^{C_{\mathrm{1}}}{\left(\psi \left(n,k\right)\right)}^{C_{\mathrm{2}}}}},                                                            
\end{equation} 
where $k\mathrm{\notin }{\mathrm{S}}_{{\mathcal{T}}_i}$, $C_{\mathrm{1}}$, and $C_{\mathrm{2}}$ are weighting parameters to balance the relevance between inverse entanglement gradient function $\theta \left(\mathrm{\cdot }\right)$ and distance function $\psi \left(\mathrm{\cdot }\right)$. 

 The remaining quantities of \eqref{ZEqnNum252838} are evaluated as 
\begin{equation} \label{43)} 
{\theta }^k_{E_{{\mathrm{L}}_l}\left(n,k\right)}\mathrm{=}\frac{\mathrm{1}}{{\mathcal{G}}'^k_{A,n}},                                                                           
\end{equation} 
and
\begin{equation} \label{44)} 
\psi \left(n,k\right)\mathrm{=P}{\mathrm{r}}^n_{E_{{\mathrm{L}}_l}\left(n,k\right)}\left|\mathbb{E}\left({\mathcal{G}}'^n_{{\mathcal{P}}_i}\right)\mathrm{-}\mathbb{E}\left({\mathcal{G}}'^k_{{\mathcal{P}}_i}\right)\right|.                                                 
\end{equation} 
 
\subsection{Algorithm}

A brief description of the entanglement-gradient routing algorithm ${\mathcal{A}}_{\mathcal{G}}$ for finding the shortest path via the entanglement gradient is as follows, see Algorithm 1. 
\setcounter{algocf}{0}
\begin{algorithm}
\DontPrintSemicolon    
\caption{\textit{Entanglement-gradient routing}}
\textbf{Step 1.} Set $t$ and ${\ell }_{\mathcal{T}}$ for threads ${\mathcal{T}}_{\mathrm{1}}\mathrm{,\dots ,}{\mathcal{T}}_t$, where ${\ell }_{\mathcal{T}}$ is a thread-threshold that limits the maximal number of nodes to be explored to ${\ell }_{\mathcal{T}}$ by a given ${\mathcal{T}}_i$. For a given ${\mathcal{T}}_i$, let ${\mathrm{S}}_{{\mathcal{T}}_i}$ be a set of already visited nodes.

\textbf{Step 2.} For a thread ${\mathcal{T}}_i$, given node $n$ and next node $z$, determine $p_{{\mathcal{T}}_i}\left(n,z\right)$ via \eqref{ZEqnNum706918}. If $z\mathrm{\notin }{\mathrm{S}}_{{\mathcal{T}}_i}$, set $p_{{\mathcal{T}}_i}\left(n,z\right)\mathrm{=0}$; otherwise, calculate $\mathrm{P}{\mathrm{r}}_{{\mathcal{T}}_i}\left(n,z\right)$ via \eqref{ZEqnNum252838}. 

\textbf{Step 3.} As a next node $n\mathrm{+1}$ is determined, update inverse link entanglement gradient ${\theta }^{n\mathrm{+1}}_{E_{{\mathrm{L}}_l}\left(n,n\mathrm{+1}\right)}$ via \eqref{ZEqnNum356733}. 

\textbf{Step 4.} Apply steps 1--3 for all threads, ${\mathcal{T}}_{\mathrm{1}}\mathrm{,\dots ,}{\mathcal{T}}_t$. 
\end{algorithm}

\subsubsection{Discussion}
In ${\mathcal{A}}_{\mathcal{G}}$, any thread ${\mathcal{T}}_i$ at a given step selects that node for which the entanglement gradient is high, i.e., the inverse link entanglement gradient of the entangled connection is low. When the inverse link entanglement gradient is high, the threads choose a different direction and entangled links. The thread threshold ${\ell }_{\mathcal{T}}$ allows for focus on a particular subset of the network for an optimal parallel realization. The distance function of \eqref{ZEqnNum867276} takes into consideration not just the absolute entanglement-gradient difference but also the inverse of the probability of the selection of a given entangled link. The threads also change their behavior as the entanglement-gradients evolve in the network, which yields dynamically changing adaptive searching. 

 For a given thread ${\mathcal{T}}_i$, the $C_{\mathrm{1}}$ and $C_{\mathrm{2}}$ weighting coefficients are crucial in the probability function of \eqref{ZEqnNum252838}  for determining the local behavior of a given thread (e.g., the routing is decentralized). The selection method of these weights is discussed next.
 
\subsection{Computational Complexity}
The computational complexity of the entanglement-gradient routing at $\left|N\right|$ nodes, with $t$ parallel threads and  a thread-threshold ${\ell }_{\mathcal{T}}$, is at most
\begin{equation} \label{ZEqnNum596070} 
\mathcal{O}\left(\left|N\right|t{\ell }_{\mathcal{T}}\right).                                                                                  
\end{equation} 
The result of \eqref{ZEqnNum596070} can be verified easily, since the maximal number of nodes visited by a given thread ${\mathcal{T}}_i$ is at most ${\ell }_{\mathcal{T}}$.
 
\section{Numerical Evidence}
\label{sec5}
In this section, we analyze the performance metrics of the link and path selection phases and the entanglement-gradient routing. 
 
\subsection{Link and Path Metrics}
\begin{figure*}[h!]
\vspace{-0.6cm}
 \begin{center}
 	 \includegraphics[angle = 0,width=1\linewidth]{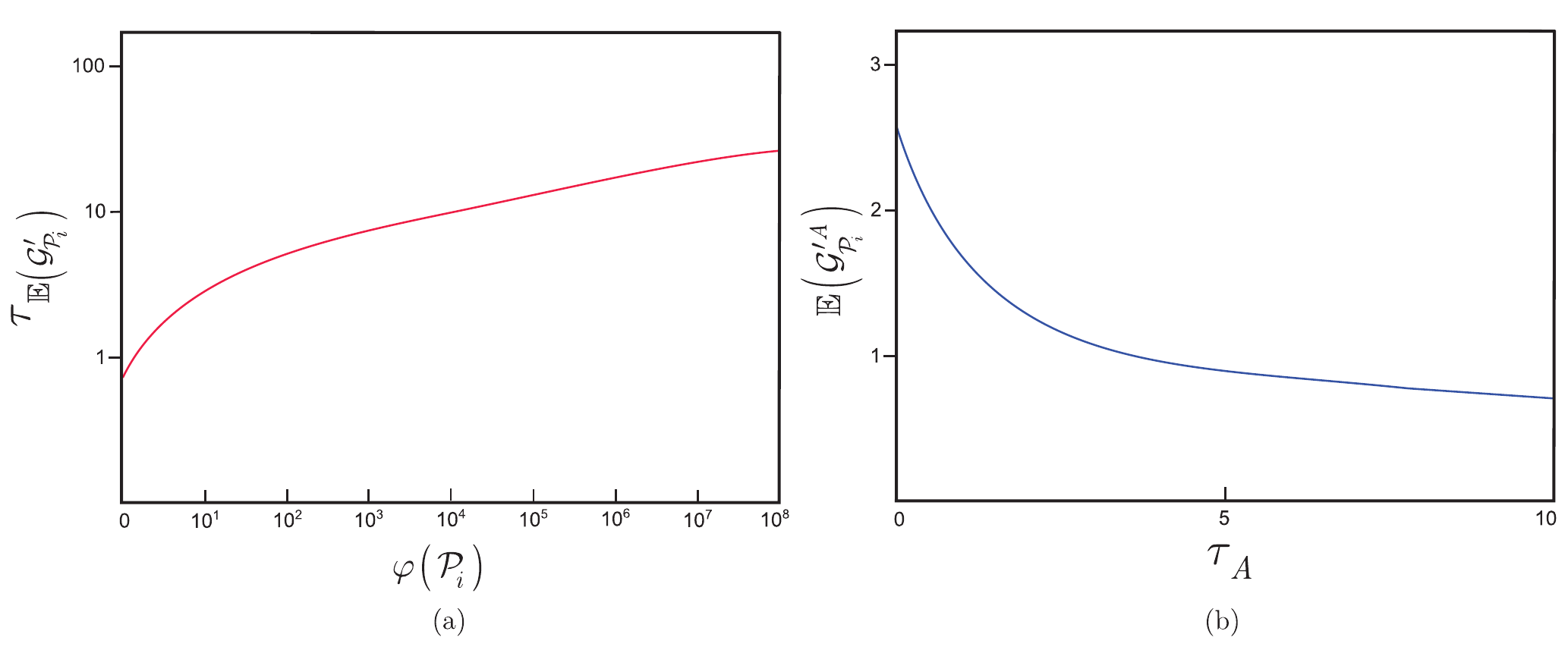}

\caption{ \textbf{(a)}: The decay rate ${\tau }_{\mathbb{E}\left({\mathcal{G}}'_{{\mathcal{P}}_i}\right)}$ (log-scale) of entanglement gradient $\mathbb{E}\left({\mathcal{G}}'_{{\mathcal{P}}_i}\right)$ for path ${\mathcal{P}}_i$, $\varphi \left({\mathcal{P}}_i\right)\mathrm{=1}0^0\mathrm{,\dots ,1}0^{\mathrm{8}}$, and $\mathrm{\partial }\mathrm{=1}$, $\mathbb{E}\left({\mathcal{G}}'_{{\mathcal{P}}_i}\right)\mathrm{=2}$. \textbf{(b)}: The $\mathbb{E}\left({\mathcal{G}}'^A_{{\mathcal{P}}_i}\right)$ of a path ${\mathcal{P}}_i$ at node $A$ as a function of ${\tau }_A$ at ${\kappa }_{AB}\mathrm{=4,}{\kappa }_B\mathrm{=2}$ and ${\mu }^A_{{\mathcal{P}}_i}\mathrm{=1}$.}
\label{fig3}
\end{center}
\end{figure*}

In this section, the proposed link and path metrics are analyzed. The decay rate ${\tau }_{\mathbb{E}\left({\mathcal{G}}'_{{\mathcal{P}}_i}\right)}$ (\eqref{ZEqnNum763249}) of entanglement gradient $\mathbb{E}\left({\mathcal{G}}'_{{\mathcal{P}}_i}\right)$ for various $\varphi \left({\mathcal{P}}_i\right)$ at ${\mathrm{\partial }}_{\mathbb{E}\left({\mathcal{G}}'_{{\mathcal{P}}_i}\right)}\mathrm{=1}$ and $\mathbb{E}\left({\mathcal{G}}'_{{\mathcal{P}}_i}\right)\mathrm{=2}$ is depicted in \fref{fig3}(a). The ${\tau }_{\mathbb{E}\left({\mathcal{G}}'_{{\mathcal{P}}_i}\right)}$ decay rate of entanglement gradient increases with the $\varphi \left({\mathcal{P}}_i\right)$ parameter of the given path ${\mathcal{P}}_i$. As the $B_F\left({\mathcal{P}}_i\right)$ entanglement throughput of that path ${\mathcal{P}}_i$ significantly deviates from the expected average ${\tilde{B}}_F\left({\mathcal{P}}_i\right)$, the entanglement gradient of ${\mathcal{P}}_i$ decreases more significantly. 
In \fref{fig3}(b), the $\mathbb{E}\left({\mathcal{G}}'^A_{{\mathcal{P}}_i}\right)$ (see \eqref{ZEqnNum406725}) of a particular path ${\mathcal{P}}_i$ at node $A$, as a function of ${\tau }_A$ at ${\kappa }_{AB}\mathrm{=4,}{\kappa }_B\mathrm{=2}$ and ${\mu }^A_{{\mathcal{P}}_i}\mathrm{=1}$, is depicted. \\

Without loss of generality, at a given ${\kappa }_n$ in a node $n$, let ${\nu }_n$ be defined as  
\begin{equation} \label{46)} 
\frac{\mathrm{-}\mathrm{2}\pi }{{\kappa }_n}\mathrm{\le }{\nu }_n\mathrm{\le }\frac{\mathrm{2}\pi }{{\kappa }_n}.                                                                             
\end{equation} 
Then, rewrite $\mathbb{E}\left({\mathcal{G}}'^n_{{\mathcal{P}}_i}\left({\nu }_n\right)\right)$ as
\begin{equation} \label{47)} 
\mathbb{E}\left({\mathcal{G}}'^n_{{\mathcal{P}}_i}\left({\nu }_n\right)\right)\mathrm{=}{\mu }^n_{{\mathcal{P}}_i}\varsigma \left({\gamma }_n\right),                                                                      
\end{equation} 
where $\varsigma \left({\gamma }_n\right)$ is a peak of function $\rho \left({\nu }_n\right)$ and $\rho \left({\nu }_n\right)$ is
\begin{equation} \label{ZEqnNum511294} 
\rho \left({\nu }_n\right)\mathrm{=}\frac{\mathrm{1}}{\sqrt{\mathrm{1+}{\gamma }^{\mathrm{2}}_n\mathrm{-}\mathrm{2}{\gamma }_n\mathrm{cos}\left({\nu }_n\right)}},                                                                 
\end{equation} 
where ${\gamma }_n$ is
\begin{equation} \label{49)} 
{\gamma }_n\mathrm{=}\frac{{\kappa }_n}{{\kappa }_n\mathrm{+}{\tau }_n}\mathrm{=}{\left(\mathrm{1+}\frac{\tau }{{\kappa }_n}\right)}^{\mathrm{-}\mathrm{1}},                                                                   
\end{equation} 
where ${\kappa }_n$ is the observation rate in node $n$and ${\tau }_n$ is the decay rate of the entanglement gradient in node $n$.

\begin{figure*}[h!]
\vspace{-0.6cm}
 \begin{center}
 	 \includegraphics[angle = 0,width=1\linewidth]{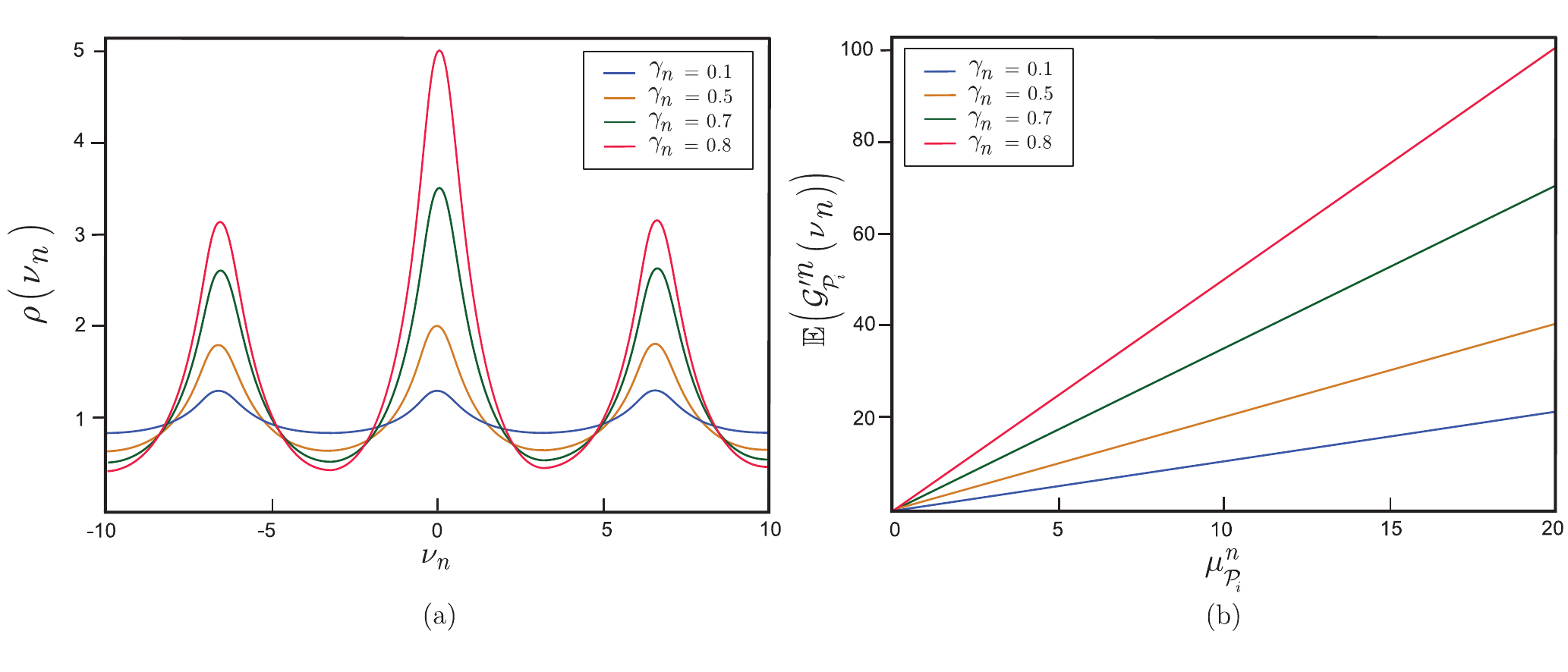}

\caption{(a): The values of $\rho \left({\nu }_n\right)$ as a function of ${\nu }_n$ for ${\gamma }_n\mathrm{=0.1,0.5,0.8,0.9}$ at node $n$. The $\varsigma \left({\gamma }_n\right)$ peak of $\rho \left({\nu }_n\right)$ at a given ${\gamma }_n$ is $\varsigma \left({\gamma }_n\right)\mathrm{=}{\mathrm{1}}/{\left(\mathrm{1-}{\gamma }_n\right)}$. (b): The mean $\mathbb{E}\left({\mathcal{G}}'^n_{{\mathcal{P}}_i}\left({\nu }_n\right)\right)\mathrm{=}{\mu }^n_{{\mathcal{P}}_i}\varsigma \left({\gamma }_n\right)$ entanglement gradient received from path ${\mathcal{P}}_i$ as a function of ${\mu }^n_{{\mathcal{P}}_i}$ for a ${\gamma }_n\mathrm{=0.1,0.5,0.8,0.9}$.}
\label{fig4}
\end{center}
\end{figure*}

 The quantity of $\varsigma \left({\gamma }_n\right)$ is derived as follows. The formula of $\rho \left({\nu }_n\right)$ (see \eqref{ZEqnNum511294}) can be rewritten as a magnitude 
\begin{equation} \label{50)} 
\rho \left({\nu }_n\right)\mathrm{=}\left|\mathcal{F}\left({\nu }_n\right)\right|,                                                                               
\end{equation} 
where $\mathcal{F}\left({\nu }_n\right)$ is defined as
\begin{equation} \label{51)} 
\mathcal{F}\left({\nu }_n\right)\mathrm{=}\frac{\mathrm{1}}{\mathrm{1-}{\gamma }_ne^{\mathrm{-}i{\nu }_n}}.                                                                           
\end{equation} 
Thus, the peak $\varsigma \left({\gamma }_n\right)$ of $\rho \left({\nu }_n\right)$ at a given ${\gamma }_n$ is yielded as 
\begin{equation} \label{ZEqnNum574939} 
\varsigma \left({\gamma }_n\right)\mathrm{=}\frac{\mathrm{1}}{\left(\mathrm{1-}{\gamma }_n\right)}.                                                                                
\end{equation} 
At a given ${\gamma }_n$ and mean ${\mu }^n_{{\mathcal{P}}_i}$ (average value of received entanglement gradient) at node $n$ for path ${\mathcal{P}}_i$, the mean of the received entanglement gradient can be rewritten from $\varsigma \left({\gamma }_n\right)$ (see \eqref{ZEqnNum574939}) as
\begin{equation} \label{53)} 
\mathbb{E}\left({\mathcal{G}}'^n_{{\mathcal{P}}_i}\left({\nu }_n\right)\right)\mathrm{=}{\mu }^n_{{\mathcal{P}}_i}\varsigma \left({\gamma }_n\right)\mathrm{=}\frac{{\mu }^n_{{\mathcal{P}}_i}}{\left(\mathrm{1-}{\gamma }_n\right)}.                                                         
\end{equation} 
The value of $\rho \left({\nu }_n\right)$ as a function of ${\nu }_n$ for various ${\gamma }_n$ is depicted in \fref{fig4}(a). The resulting mean entanglement gradient $\mathbb{E}\left({\mathcal{G}}'^n_{{\mathcal{P}}_i}\left({\nu }_n\right)\right)$, as a function of ${\mu }^n_{{\mathcal{P}}_i}$ for various ${\gamma }_n$ of path ${\mathcal{P}}_i$ at node $n$, is depicted in \fref{fig4}(b). As the ${\mu }^n_{{\mathcal{P}}_i}$ average value of the received entanglement gradient increases, the mean entanglement gradient $\mathbb{E}\left({\mathcal{G}}'^n_{{\mathcal{P}}_i}\left({\nu }_n\right)\right)$ becomes more significant, specifically for high values of ${\gamma }_n$.\\

At a $\mathrm{0}\mathrm{\le }\mathrm{\Pi }\mathrm{\le }\mathrm{1}$ tuning parameter (a fraction of peak value), let ${\kappa }^{\mathrm{*}}_n$ be a cutoff observation rate (critical received $d$-dimensional entangled states per sec) defined at a given observation rate ${\kappa }_n$ as
\begin{equation} \label{54)} 
{\kappa }^{\mathrm{*}}_n\mathrm{=}\frac{{\kappa }_n}{\mathrm{2}\pi }\mathrm{co}{\mathrm{s}}^{\mathrm{-}\mathrm{1}}\left(\frac{{\kappa }^{\mathrm{2}}_n\mathrm{+}{\kappa }_n{\tau }_n\mathrm{-}\frac{{\tau }^{\mathrm{2}}_n\mathrm{+}{\mathrm{\Pi }}^{\mathrm{2}}{\tau }^{\mathrm{2}}_n}{\mathrm{2}{\mathrm{\Pi }}^{\mathrm{2}}}}{{\kappa }^{\mathrm{2}}_n\mathrm{+}{\kappa }_n{\tau }_n}\right). 
\end{equation} 
If ${\kappa }_n\mathrm{>}{\kappa }^{\mathrm{*}}_n$ at a given ${\tau }_n$ and $\mathrm{\Pi }$, then in node $n$, the value of the total received entanglement gradient will not adapt to the actually received total value of entanglement gradients, i.e., ${\kappa }^{\mathrm{*}}_n$ serves a cutoff in the observation rate.  

\begin{figure*}[h!]
\vspace{-0.6cm}
 \begin{center}
 	 \includegraphics[angle = 0,width=1\linewidth]{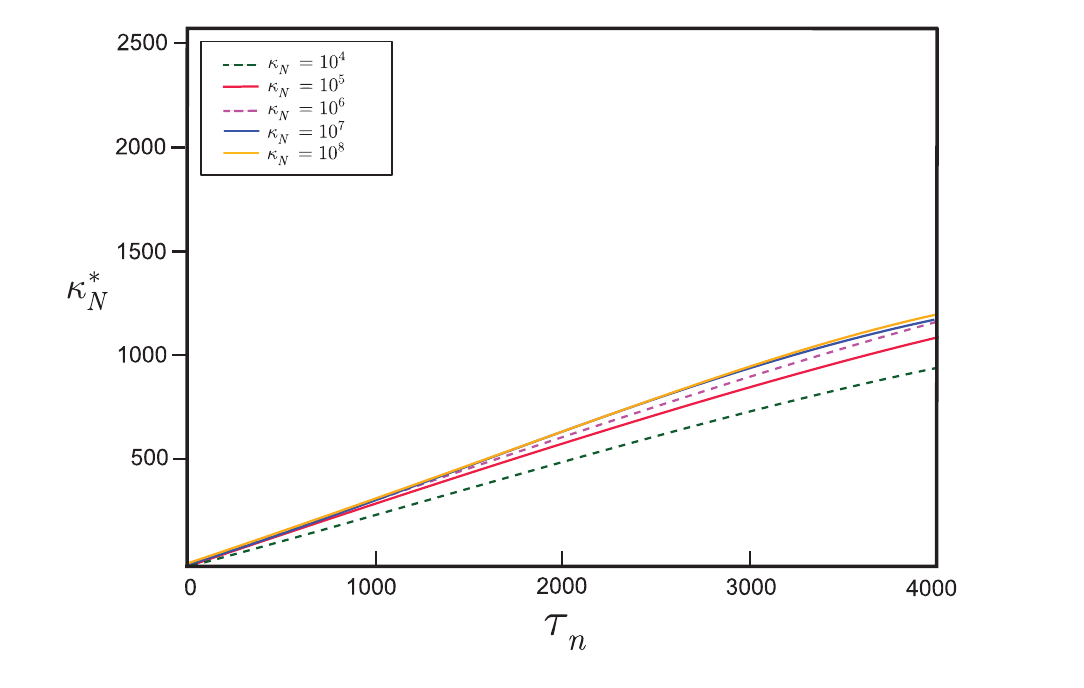}

\caption{The values of ${\kappa }^{\mathrm{*}}_n$ as a function of ${\tau }_n$ for ${\kappa }_n\mathrm{=1}0^{\mathrm{4}}\mathrm{,1}0^{\mathrm{5}}\mathrm{,1}0^{\mathrm{6}}\mathrm{,1}0^{\mathrm{7}}\mathrm{,1}0^{\mathrm{8}}$ at $\mathrm{\Pi }\mathrm{=0.5}$.}
\label{fig5}
\end{center}
\end{figure*}

 The values of ${\kappa }^{\mathrm{*}}_n$ as a function of ${\tau }_n$ for various ${\kappa }_n$ at $\mathrm{\Pi }\mathrm{=0.5}$ (in analogue to a $\mathrm{-}\mathrm{3}$ dB cutoff [\cref{ref15}-\cref{ref18}] are shown in \fref{fig5}. The ${\kappa }^{\mathrm{*}}_n$ cutoff observation rate is controllable by ${\tau }_n$ and the impact of an actual ${\kappa }_n$ rate on ${\kappa }^{\mathrm{*}}_n$ is almost negligible.\\

\subsection{Decentralized Routing}

 The routing procedure is discussed by the $\mathrm{P}{\mathrm{r}}_{{\mathcal{T}}_i}\left(n,z\right)$ probability function of \eqref{ZEqnNum252838}. In \eqref{ZEqnNum252838}, the  $C_{\mathrm{1}}$ and $C_{\mathrm{2}}$ weights have a crucial role and are determined as follows. 

 If the average value ${\mu }^n\left({\mathrm{\Phi }}^{s,n}_{\mathcal{P}}\right)$ (see \eqref{ZEqnNum751770}) is low, then $C_{\mathrm{1}}$ is high and $C_{\mathrm{2}}$ is low. In this case an another way and a different node but not $z$ is selected at an actual node $n$. For a high ${\mu }^n\left({\mathrm{\Phi }}^{s,n}_{\mathcal{P}}\right)$, $C_{\mathrm{2}}$ picks up a high value and $C_{\mathrm{1}}$ is low. In this case the current target node $z$ is selected at node $n$. 

Thus, for a given threshold $o$ on ${\mu }^n\left({\mathrm{\Phi }}^{s,n}_{\mathcal{P}}\right)$, ${\mu }^n\left({\mathrm{\Phi }}^{s,n}_{\mathcal{P}}\right)\mathrm{\le }o$, the selection rule for weights $\left\{C_{\mathrm{1}},C_{\mathrm{2}}\right\}$ is 
\begin{equation} \label{ZEqnNum480384} 
\left\{C_{\mathrm{1}},C_{\mathrm{2}}\right\}\mathrm{=}\left\{ \begin{array}{l}
C_{\mathrm{1}}\mathrm{\to }\mathrm{1,}C_{\mathrm{2}}\mathrm{\to }\mathrm{0,}\textnormal{ if }{\mu }^n\left({\mathrm{\Phi }}^{s,n}_{\mathcal{P}}\right)\mathrm{\le }o, \\ 
C_{\mathrm{1}}\mathrm{\to }\mathrm{0,}C_{\mathrm{2}}\mathrm{\to }\mathrm{1,otherwise.} \end{array}
\right. 
\end{equation} 
As a corollary of \eqref{ZEqnNum480384}, a high value of $C_{\mathrm{1}}$ and a low value of $C_{\mathrm{2}}$ increases the network area to be explored by a thread, while for a low value of $C_{\mathrm{1}}$ and a high value of $C_{\mathrm{2}}$, the number of explored nodes is smaller. 

\begin{figure*}[h!]
\vspace{-0.6cm}
 \begin{center}
 	 \includegraphics[angle = 0,width=1\linewidth]{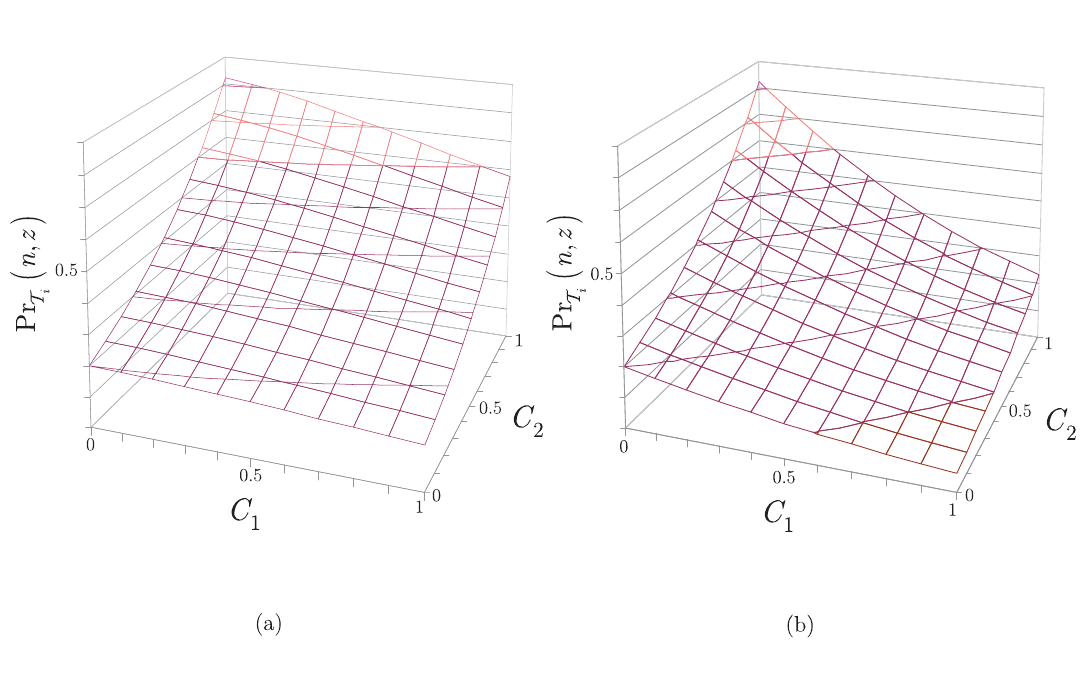}

\caption{(a):The $\mathrm{P}{\mathrm{r}}_{{\mathcal{T}}_i}\left(n,z\right)$ probability of selecting a next node $z$ from a current node $n$ for a given path ${\mathcal{P}}_i$, $\mathrm{0}\mathrm{\le }C_{\mathrm{1}}\mathrm{\le }\mathrm{1}$ and $\mathrm{0}\mathrm{\le }C_{\mathrm{2}}\mathrm{\le }\mathrm{1}$, at a setting of ${\theta }^z_{E_{{\mathrm{L}}_l}\left(n,z\right)}\mathrm{=0.5}$, $\psi \left(n,z\right)\mathrm{=5}$, $k\mathrm{=5}$, and at (b): ${\theta }^{z}_{{E}_{{\mathrm{L}}_{l}}\left(n,z\right)}\mathrm{=0.2}$, $\psi \left(n,z\right)\mathrm{=5}$, $k\mathrm{=5}$.}
\label{fig6}
\end{center}
\end{figure*}

 The values of $\mathrm{P}{\mathrm{r}}_{{\mathcal{T}}_i}\left(n,z\right)$ \eqref{ZEqnNum252838} as a function of weight coefficients $C_{\mathrm{1}}$ and $C_{\mathrm{2}}$ for a given path ${\mathcal{P}}_i$, current node $n$, and next node $z$ at a particular thread ${\mathcal{T}}_i$ and network setting are depicted in \fref{fig6}. In \fref{fig6}(a), the inverse link entanglement gradient is ${\theta }^z_{E_{{\mathrm{L}}_l}\left(n,z\right)}\mathrm{=0.5}$, while in \fref{fig6}(b), it has the value of ${\theta }^z_{E_{{\mathrm{L}}_l}\left(n,z\right)}\mathrm{=0.2}$. 

\subsection{Achievable Entanglement Fidelity in the Protocol}

 Assuming an ideal recovery operation $\mathcal{R}$ with an optimal quantum error correction [\cref{ref20}] in the proposed routing mechanism, the $F$ entanglement fidelity is evaluated as 
\begin{equation} \label{ZEqnNum896752} 
F\mathrm{=}\left\langle \widetilde{\mathrm{\Psi }}\mathrel{\left|\vphantom{\widetilde{\mathrm{\Psi }} \left.\mathcal{R}\left({\rho }_f\right)\right|\widetilde{\mathrm{\Psi }}}\right.\kern-\nulldelimiterspace}\left.\mathcal{R}\left({\rho }_f\right)\right|\widetilde{\mathrm{\Psi }}\right\rangle ,                                                                                      
\end{equation} 
where $\widetilde{\mathrm{\Psi }}$ is a shared Bell pair between the final stations, while ${\rho }_f$ is the input density matrix of $\mathcal{R}$. 

\begin{figure*}[h!]
\vspace{-0.6cm}
 \begin{center}
 	 \includegraphics[angle = 0,width=1\linewidth]{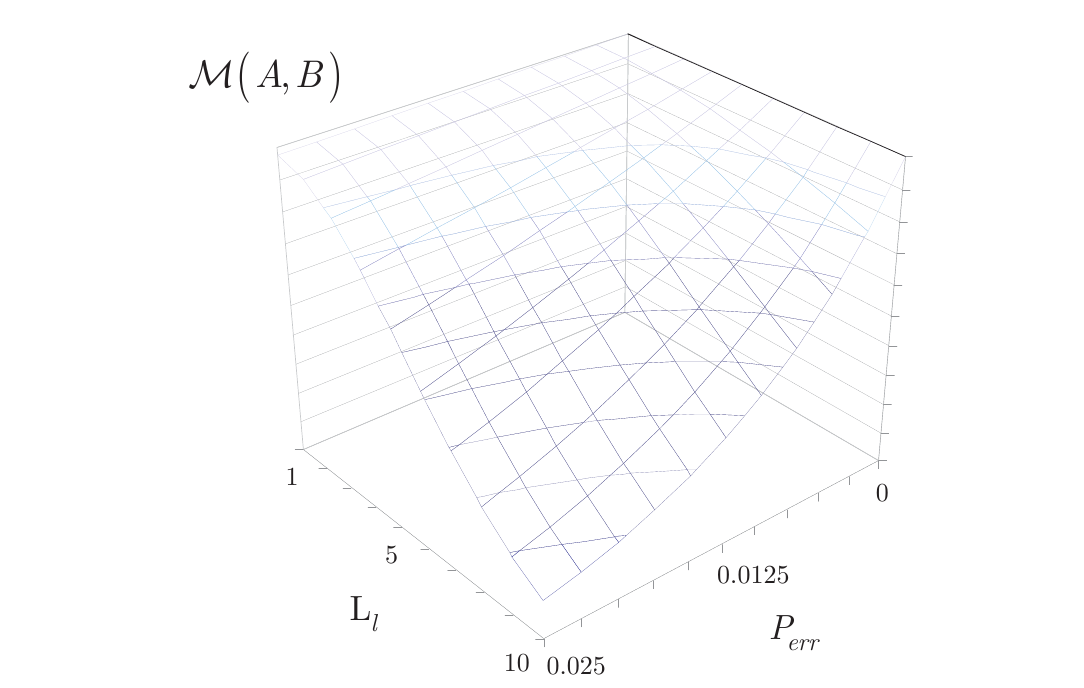}

\caption{The $F$ entanglement fidelity as obtainable from the correlation measurement $\mathcal{M}\left(A,B\right)\mathrm{\approx }\sqrt{F}$ between final stations $A$ and $B$, at per-node error probability $P_{err}\mathrm{=}\left[\mathrm{0,0.025}\right]$, hop distance $d{\left(A,B\right)}_{{\mathrm{L}}_l}\mathrm{=}{\mathrm{2}}^{l\mathrm{-}\mathrm{1}}$, and entanglement level ${\mathrm{L}}_l\mathrm{=}\left[\mathrm{1,10}\right]$.}
\label{fig7}
\end{center}
\end{figure*}

 For an ${\mathrm{L}}_l$-level entangled link $E_{{\mathrm{L}}_l}\left(A,B\right)$ with hop-distance $d{\left(A,B\right)}_{{\mathrm{L}}_l}\mathrm{=}{\mathrm{2}}^{l\mathrm{-}\mathrm{1}}$ between final stations $A$ and $B$  and per-node error probability $P_{err}$ (that includes the effective logical error probability $Q$ and other residual errors ${\varepsilon }_{res}$ in the nodes) in the $d{\left(A,B\right)}_{{\mathrm{L}}_l}\mathrm{+1}$ total stations, after some calculations the entanglement fidelity \eqref{ZEqnNum896752} can be rewritten as [\cref{ref20}]
\begin{equation} \label{57)} 
 \begin{array}{l}
\begin{split}
F&\mathrm{=}{\left(\mathrm{1-}P_{err}\right)}^{\mathrm{2}\left(d{\left(A,B\right)}_{{\mathrm{L}}_l}\mathrm{+1}\right)\mathrm{-}\mathrm{2}} \\ 
&\mathrm{=}{\left(\mathrm{1-}\left(Q\mathrm{+}{\varepsilon }_{res}\right)\right)}^{\mathrm{2}d{\left(A,B\right)}_{{\mathrm{L}}_l}}. 
\end{split}
\end{array}
\end{equation} 
The performance of the routing is approachable by the correlation measurement $\mathcal{M}\left(A,B\right)$ between the final stations $A$ and $B$, which quantity practically yields the corresponding fidelity information as [\cref{ref20}]
\begin{equation} \label{58)} 
 \begin{array}{l}
\begin{split}
\mathcal{M}\left(A,B\right)&\mathrm{\approx }\sqrt{F} \\ 
&\mathrm{\approx }{\left(\mathrm{1-}P_{err}\right)}^{d{\left(A,B\right)}_{{\mathrm{L}}_l}\mathrm{+1}} \\ 
&\mathrm{=}{\left(\mathrm{1-}\left(Q\mathrm{+}{\varepsilon }_{res}\right)\right)}^{{\mathrm{2}}^{l\mathrm{-}\mathrm{1}}\mathrm{+1}}. 
\end{split}
\end{array}
\end{equation} 
The results for $\mathcal{M}\left(A,B\right)$ in function of per-node error probability $P_{err}$, and level ${\mathrm{L}}_l$ of  the entangled link $E_{{\mathrm{L}}_l}\left(A,B\right)$ between the final stations $A$ and $B$  are depicted in \fref{fig7}. 

 In the protocol, the $F$ entanglement fidelity of the final Bell pair $\widetilde{\mathrm{\Psi }}$ between $A$ and $B$ (see \eqref{57)}) achieves a theoretical maximum at $d{\left(A,B\right)}_{{\mathrm{L}}_l}\mathrm{-}\mathrm{1}$ intermediate quantum repeaters, and at operators $\mathcal{R}$ and $\mathcal{C}$ [\cref{ref19}, \cref{ref47}]. Using $P_{err}$, the $P^{tot}_{succ}\left(\mathcal{R}\right)$ total success probability of the recovery operation $\mathcal{R}$ for the intermediate nodes is evaluated as
\begin{equation} \label{ZEqnNum315218} 
P^{tot}_{succ}\left(\mathcal{R}\right)\mathrm{=}{\left(\mathrm{1-}P_{err}\right)}^{\mathrm{2}\left(\left(d{\left(A,B\right)}_{{\mathrm{L}}_l}\mathrm{+1}\right)\mathrm{-}\mathrm{2}\right)}\mathrm{=}{\left(\mathrm{1-}\left(Q\mathrm{+}{\varepsilon }_{res}\right)\right)}^{\mathrm{2}\left(d{\left(A,B\right)}_{{\mathrm{L}}_l}\mathrm{-}\mathrm{1}\right)}, 
\end{equation} 
while the internal error-correction operation $\mathcal{C}$ has a $P^{\widetilde{\mathrm{\Psi }}}_{succ}\left(\mathcal{C}\right)$ success probability with respect to the final state $\widetilde{\mathrm{\Psi }}$ as
\begin{equation} \label{ZEqnNum698665} 
P^{\widetilde{\mathrm{\Psi }}}_{succ}\left(\mathcal{C}\right)\mathrm{=}{\left(\mathrm{1-}\left(Q\mathrm{+}{\varepsilon }_{res}\right)\right)}^{\mathrm{2}}. 
\end{equation} 
The success probabilities in \eqref{ZEqnNum315218} and \eqref{ZEqnNum698665} yield an estimation [\cref{ref19}, \cref{ref47}] for the entanglement fidelity $F$ of $\widetilde{\mathrm{\Psi }}$ as
\begin{equation} \label{61)} 
F\mathrm{\approx }P^{tot}_{succ}\left(\mathcal{R}\right)P^{\widetilde{\mathrm{\Psi }}}_{succ}\left(\mathcal{C}\right), 
\end{equation} 
which therefore practically yields \eqref{57)}.

\subsection{Security of the Protocol}
Based on the $F$ entanglement fidelity \eqref{57)} of $\widetilde{\mathrm{\Psi }}$, the security of the protocol can be characterized [\cref{ref19}, \cref{ref47}] as follows. At a particular final key $K$ (a shared bitstring between $A$ and $B$), the proposed protocol guarantees that the maximum information leaked to an eavesdropper $E$ (Eve) is upper bounded [\cref{ref47}] as
\begin{equation} \label{62)} 
I\mathrm{(}E\mathrm{:}K\mathrm{)}\mathrm{\le }{\mathrm{2}}^{\mathrm{-}c}\mathrm{+}{\mathrm{2}}^{\mathcal{O}\left(\mathrm{-}\mathrm{2}s\right)},                                                                             
\end{equation} 
where $I\mathrm{(}E\mathrm{:}K\mathrm{)}$ is the mutual information of Eve, and 
\begin{equation} \label{63)} 
c\mathrm{=}s\mathrm{-}\mathrm{lo}{\mathrm{g}}_{\mathrm{2}}\left(\mathrm{2+}s\mathrm{+}\frac{\mathrm{1}}{\mathrm{ln2}}\right), 
\end{equation} 
while $s$ is evaluated as
\begin{equation} \label{64)} 
s\mathrm{=-lo}{\mathrm{g}}_{\mathrm{2}}\left(\mathrm{1-}F\right),                                                                                  
\end{equation} 
at a particular entanglement fidelity $F$ \eqref{57)} of $\widetilde{\mathrm{\Psi }}$. 

The protocol therefore also provides a practical framework to realize quantum key distribution over long distances. 

\section{Conclusions}
\label{sec6}
In this work, we defined the entanglement-gradient routing method for quantum repeater networks. The routing scheme is based on the fundamentals of swarm intelligence in order to find the optimal shortest path in entangled quantum networks. We defined the terms of entanglement utility and link and path entanglement gradient, and proposed the routing metrics. The routing metrics are derived from the characteristics of entangled links, entanglement throughput capabilities, and the distribution of the entangled states. The method allows for moderate complexity routing in quantum repeater networks by fusing the relevant characteristics of entanglement distribution and swarm intelligence theory. The scheme can be directly applied in quantum networking, future quantum Internet, and experimental long-distance quantum communications. 

\section*{Acknowledgements}
This work was partially supported by the Hungarian Scientific Research Fund - OTKA K-112125, and by the COST Action MP1006.
 
\section*{References}
\begin{enumerate}[ {[}1{]} ]
\item \label{ref1} Van Meter, R. \textit{Quantum Networking}. ISBN 1118648927, 9781118648926, John Wiley and Sons Ltd (2014).

\item \label{ref2} Imre, S., Gyongyosi, L. \textit{Advanced Quantum Communications - An Engineering Approach}. New Jersey, Wiley-IEEE Press (2012). 

\item \label{ref3} Van Meter, R., Satoh, T., Ladd, T. D., Munro, W. J. and Nemoto, K. Path Selection for Quantum Repeater Networks. \textit{Networking Science}, 3, 82-95 (2013).

\item \label{ref4} Van Meter, R., Ladd, T. D., Munro, W. J. and Nemoto, K. System Design for a Long-Line Quantum Repeater. \textit{IEEE/ACM Transactions on Networking}, 17, 1002-1013 (2009).

\item \label{ref5} Lloyd, S., Shapiro, J. H., Wong, F. N. C., Kumar, P., Shahriar, S. M. and Yuen, H. P. Infrastructure for the quantum Internet. \textit{ACM SIGCOMM Computer Communication Review}, 34, 9--20 (2004).

\item \label{ref6} Hanzo, L., Haas, H., Imre, S., O'Brien, D., Rupp, M. and Gyongyosi, L. Wireless Myths, Realities, and Futures: From 3G/4G to Optical and Quantum Wireless. \textit{Proceedings of the IEEE}, 100, 1853-1888 (2012).

\item \label{ref7} Lloyd, S., Mohseni, M. and Rebentrost, P. Quantum principal component analysis. \textit{Nature Physics}, 10, 631 (2014).

\item \label{ref8} Lloyd, S. Capacity of the noisy quantum channel. \textit{Physical Rev. A}, 55:1613--1622 (1997).

\item \label{ref9} Gyongyosi, L. Quantum Imaging of High-Dimensional Hilbert Spaces with Radon Transform. \textit{International Journal of Circuit Theory and Applications}, Wiley. DOI: 10.1002/cta.2332 (2017).

\item \label{ref10} Yuan, Z., Chen, Y., Zhao, B., Chen, S., Schmiedmayer. J. and Pan, J. W. Experimental demonstration of a BDCZ quantum repeater node. \textit{Nature}, 454, 1098-1101 (2008).

\item \label{ref11} Di Caro, G., Ducatelle, F. and Gambardella, L. M. AdHocNet: An Adaptive Nature-Inspired Algorithm for Routing in placeMobile Ad-Hoc Networks. \textit{European Transactions on Telecommunications}, 16, 443--455 (2005).

\item \label{ref12} Bonabeau, E., Dorigo, M. and Theraulaz, G. \textit{Swarm Intelligence: From Natural to Artificial Systems}. ISBN 0-19-513159-2, Oxford University Press (1999).

\item \label{ref13} Meuleau, N., Dorigo, M. Ant Colony Optimization and Stochastic Gradient Descent. \textit{Artificial Life}, 8, 103-121 (2002).

\item \label{ref14} Roth, M., Wicker, S. Termite Ad-hoc Networking with Stigmergy. \textit{The Second Mediterranean Workshop on Ad-Hoc Networks}. 2937-2941, DOI: 10.1109/GLOCOM.2003.1258772 (2003).

\item \label{ref15} Roth, M., Wicker, S. Performance Evaluation of Pheromone Update in Swarm Intelligent MANETs. \textit{Mobile and Wireless Communication Networks}, 335-346, DOI: 10.1007/0-387-23150-1\_29 (2005).

\item \label{ref16} Simone, G., Gadia, D., Farup, and Rizzi, A. Ant Colony for Locality Foraging in Image Enhancement. \textit{Multiobjective Swarm Intelligence}(Editors: Dehuri, S., Jagadev, A. K. and Panda, M). Springer, 123-142 (2015).

\item \label{ref17} Roth, M., Wicker, S. Asymptotic Pheromone Behavior in Swarm Intelligent MANETs: An Analytical Analysis of Routing Behavior. \textit{Mobile and Wireless Communications Networks}. DOI: 10.1007/0-387-23150-1\_29, 335-346 (2004).

\item \label{ref18} Roth, M. \textit{Termite: A Swarm Intelligent Routing Algorithm for placeMobile Wireless Ad-Hoc Networks}. PhD Thesis, Cornell University (2005).

\item \label{ref19} Jiang, L., Taylor, J. M., Nemoto, K., Munro, W. J., Van Meter, R. and Lukin, M. D. Quantum repeater with encoding. \textit{Phys. Rev. A}, 79:032325 (2009).

\item \label{ref20} Xiao, Y. F., Gong, Q. Optical microcavity: from fundamental physics to functional photonics devices. \textit{Science Bulletin}, 61, 185-186 (2016).

\item \label{ref21} Zhang, W. et al. Quantum Secure Direct Communication with Quantum Memory. \textit{Phys. Rev. Lett}. 118, 220501 (2017).

\item \label{ref22} Biamonte, J. et al. Quantum Machine Learning. \textit{Nature}, 549, 195-202 (2017). 

\item \label{ref23} Lloyd, S. Mohseni, M. and Rebentrost, P. Quantum algorithms for supervised and unsupervised machine learning. \textit{arXiv:1307.0411} (2013).

\item \label{ref24} Dijkstra, E. W. A note on two problems in connexion with graphs. \textit{Numerische Mathematik}, 1(1): 269--271 (1959).

\item \label{ref25} Chou, C., Laurat, J., Deng, H., Choi, K. S., de Riedmatten, H., Felinto, D. and Kimble, H. J. Functional quantum nodes for entanglement distribution over scalable quantum networks. \textit{Science}, 316(5829):1316--1320 (2007).

\item \label{ref26} Kimble, H. J. The quantum Internet. \textit{Nature}, 453:1023--1030 (2008).

\item \label{ref27} Sheng, Y. B., Zhou, L. Distributed secure quantum machine learning. \textit{Science Bulletin}, 62, 1025-2019 (2017). 

\item \label{ref28} Gyongyosi, L. The Correlation Conversion Property of Quantum Channels. \textit{Quantum Information Processing}, 13, 467--473 (2014).

\item \label{ref29} Kok, P., Munro, W. J., Nemoto, K., Ralph, T. C., Dowling, J. P. and Milburn, G. J. Linear optical quantum computing with photonic qubits. \textit{Rev. Mod. Phys}. 79, 135-174 (2007).

\item \label{ref30} Gisin, N. and Thew, R. Quantum Communication. \textit{Nature Photon}. 1, 165-171 (2007).

\item \label{ref31} Enk, S. J., Cirac, J. I. and Zoller, P. Photonic channels for quantum communication. \textit{Science}, 279, 205-208 (1998).

\item \label{ref32} Briegel, H. J., Dur, W., Cirac, J. I. and Zoller, P. Quantum repeaters: the role of imperfect local operations in quantum communication. \textit{Phys. Rev. Lett.} 81, 5932-5935 (1998).

\item \label{ref33} Dur, W., Briegel, H. J., Cirac, J. I. and Zoller, P. Quantum repeaters based on entanglement purification. \textit{Phys. Rev. A, }59, 169-181 (1999).

\item \label{ref34} Duan, L. M., Lukin, M. D., Cirac, J. I. and Zoller, P. Long-distance quantum communication with atomic ensembles and linear optics. \textit{Nature,} 414, 413-418 (2001).

\item \label{ref35} Van Loock, P., Ladd, T. D., Sanaka, K., Yamaguchi, F., Nemoto, K., Munro, W. J. and Yamamoto, Y. Hybrid quantum repeater using bright coherent light. \textit{Phys. Rev. Lett, }96, 240501 (2006).

\item \label{ref36} Zhao, B., Chen, Z. B., Chen, Y. A., Schmiedmayer, J. and Pan, J. W. Robust creation of entanglement between remote memory qubits.\textit{ Phys. Rev. Lett. }98, 240502 (2007).

\item \label{ref37} Goebel, A. M., Wagenknecht, G., Zhang, Q., Chen, Y., Chen, K., Schmiedmayer, J. and Pan, J. W. Multistage Entanglement Swapping.\textit{Phys. Rev. Lett}. 101, 080403 (2008).

\item \label{ref38} Simon C., de Riedmatten H., Afzelius M., Sangouard N., Zbinden H. and Gisin N. Quantum Repeaters with Photon Pair Sources and Multimode Memories. \textit{Phys. Rev. Lett}. 98, 190503 (2007).

\item \label{ref39} Tittel, W., Afzelius, M., Chaneliere, T., Cone, R. L., Kroll, S., Moiseev, S. A. and Sellars, M. Photon-echo quantum memory in solid state systems. \textit{Laser Photon. Rev}. 4, 244-267 (2009).

\item \label{ref40} Sangouard, N., Dubessy, R. and Simon, C. Quantum repeaters based on single trapped ions. \textit{Phys. Rev. A}, 79, 042340 (2009).

\item \label{ref41} Dur, W. and Briegel, H. J. Entanglement purification and quantum error correction. \textit{Rep. Prog. Phys}, 70, 1381-1424 (2007).

\item \label{ref42} Munro, W. J., Harrison, K. A., Stephens, A. M., Devitt, S. J. and Nemoto, K. From quantum multiplexing to high-performance quantum networking. \textit{Nature Photon}. 4, 792-796 (2010).

\item \label{ref43} Sangouard, N., Simon, C., de Riedmatten, H. and Gisin, N. Quantum repeaters based on atomic ensembles and linear optics. \textit{Rev. Mod. Phys}.  83, 33-80 (2011).

\item \label{ref44} Collins, O. A., Jenkins, S. D., Kuzmich, A. and Kennedy, T. A. Multiplexed Memory-Insensitive Quantum Repeaters.\textit{Phys. Rev. Lett}. 98, 060502 (2007).

\item \label{ref45} Ralph, T. C., Hayes, A. J. F. and Gilchrist, A. Loss-Tolerant Optical Qubits. \textit{Phys. Rev. Lett}. 95, 100501 (2005).

\item \label{ref46} Kwiat, P. G. Hyper-entangled states. \textit{J. Mod. Opt}. 44, 2173-2184 (1997).

\item \label{ref47} Lo, H. K. and Chau, H. F. \textit{Science}, 283, 2050 (1999).

\item \label{ref48} Shor, P. W. Scheme for reducing decoherence in quantum computer memory.\textit{Phys. Rev. A}, 52, R2493-R2496 (1995).
\end{enumerate}
\newpage

\appendix
\setcounter{table}{0}
\setcounter{figure}{0}
\setcounter{equation}{0}
\renewcommand{\thetable}{\Alph{section}.\arabic{table}}
\renewcommand{\thefigure}{\Alph{section}.\arabic{figure}}
\renewcommand{\theequation}{\Alph{section}.\arabic{equation}}

\setlength{\arrayrulewidth}{0.1mm}
\setlength{\tabcolsep}{5pt}
\renewcommand{\arraystretch}{1.5}
\section{Appendix}
\subsection{Source-Dependent Link Selection}
\label{secA1}
The source-dependent link probability is derived as follows.

For the $A\mathrm{\to }B$ scenario the results are analogous to \eqref{ZEqnNum536222}. Therefore, utilizing ${\mathcal{G}}'^y_{z,B}$ in a current node $y$ with neighbor node $z$, the $\mathrm{P}{\mathrm{r}}^y_{E_{{\mathrm{L}}_l}\left(y,z\right)}$ probability that from node $y$ the entangled link $E_{{\mathrm{L}}_l}\left(y,z\right)$ is selected to reach destination $B$ is evaluated as 
\begin{equation} \label{ZEqnNum980976} 
\begin{split}
\mathrm{P}{\mathrm{r}}^y_{E_{{\mathrm{L}}_l}\left(y,z\right)}\\&\mathrm{=}\frac{{\left({\mathcal{G}}'^y_{z,B}\mathrm{+}\mathrm{\partial }\right)}^{\chi }}{\sum_k{{\left({\mathcal{G}}'^y_{k,B}\mathrm{+}\mathrm{\partial }\right)}^{\chi }}}\\&\mathrm{=}\frac{{\left(\left({\mathcal{G}}^y_{z,B}e^{\mathrm{-}\tau \left(\mathit{\Delta}B_F\left(E_{{\mathrm{L}}_l}\left(y,z\right)\right)\right)}\mathrm{+}{\lambda }'_{E_{{\mathrm{L}}_l}\left(y,z\right)}\right)\mathrm{+}\mathrm{\partial }\right)}^{\chi }}{\sum_k{{\left(\left({\mathcal{G}}^y_{k,B}e^{\mathrm{-}\tau \left(\mathit{\Delta}B_F\left(E_{{\mathrm{L}}_l}\left(y,k\right)\right)\right)}\right)\mathrm{+}\mathrm{\partial }\right)}^{\chi }}},                
\end{split}
\end{equation} 
where $\mathrm{\partial }\mathrm{\ge }\mathrm{0}$ is a threshold parameter, while $\chi \mathrm{\ge }\mathrm{0}$ is a tuning parameter.   

For the reverse direction $B\mathrm{\to }A$, let $\mathrm{P}{\mathrm{r}}^y_{E_{{\mathrm{L}}_l}\left(x,y\right)}$ be the probability that from node $y$ the entangled link $E_{{\mathrm{L}}_l}\left(x,y\right)$ is selected from $y$ to reach source $A$. 

The $\mathrm{P}{\mathrm{r}}^y_{E_{{\mathrm{L}}_l}\left(x,y\right)}$ entanglement gradient distribution is defined as
\begin{equation} \label{ZEqnNum775885} 
\begin{split}
\mathrm{P}{\mathrm{r}}^y_{E_{{\mathrm{L}}_l}\left(x,y\right)}\\&\mathrm{=}\frac{{\left({\mathcal{G}}'^y_{A,x}\mathrm{+}\mathrm{\partial }\right)}^{\chi }}{\sum_j{{\left({\mathcal{G}}'^y_{A,j}\mathrm{+}\mathrm{\partial }\right)}^{\chi }}}\\&\mathrm{=}\frac{{\left(\left({\mathcal{G}}^y_{A,x}e^{\mathrm{-}\tau \left(\mathit{\Delta}B_F\left(E_{{\mathrm{L}}_l}\left(x,y\right)\right)\right)}\mathrm{+}{\lambda }'_{E_{{\mathrm{L}}_l}\left(x,y\right)}\right)\mathrm{+}\mathrm{\partial }\right)}^{\chi }}{\sum_j{{\left({\mathcal{G}}^y_{A,j}e^{\mathrm{-}\tau \left(\mathit{\Delta}B_F\left(E_{{\mathrm{L}}_l}\left(j,y\right)\right)\right)}\mathrm{+}\mathrm{\partial }\right)}^{\chi }}},                 
\end{split}
\end{equation} 
where $x$ is the neighbor of $y$ with entangled link $E_{{\mathrm{L}}_l}\left(x,y\right)$ with source node $A$.

\newpage
\subsection{Normalized Distribution}

The normalized distribution $p^y_{E_{{\mathrm{L}}_l}\left(y,z\right)}$ that entangled link $E_{{\mathrm{L}}_l}\left(y,z\right)$ is selected from $y$ to reach 

From \eqref{ZEqnNum980976} and \eqref{ZEqnNum775885}, the $p^y_{E_{{\mathrm{L}}_l}\left(y,z\right)}$ normalized probability distribution that entangled link $E_{{\mathrm{L}}_l}\left(y,z\right)$ is selected from $y$ to reach destination $B$ is as
\begin{equation} \label{ZEqnNum338413} 
p^y_{E_{{\mathrm{L}}_l}\left(y,z\right)}\\\mathrm{=}\frac{\mathrm{P}{\mathrm{r}}^y_{E_{{\mathrm{L}}_l}\left(y,z\right)}{\left(\mathrm{P}{\mathrm{r}}^y_{E_{{\mathrm{L}}_l}\left(x,y\right)}\right)}^{\mathrm{-}\xi }}{\sum_k{\mathrm{P}{\mathrm{r}}^y_{E_{{\mathrm{L}}_l}\left(y,k\right)}\sum_j{{\left(\mathrm{P}{\mathrm{r}}^y_{E_{{\mathrm{L}}_l}\left(j,y\right)}\right)}^{\mathrm{-}\xi }}}},                                    
\end{equation} 
where $\xi \mathrm{\ge }\mathrm{0}$ is a weight on the source entanglement gradient, while $\mathrm{P}{\mathrm{r}}^y_{E_{{\mathrm{L}}_l}\left(y,k\right)}$ is the probability that link $E_{{\mathrm{L}}_l}\left(y,k\right)$, where $k\mathrm{\in }V\mathrm{-}x$, will be selected at node $y$, evaluated as
\begin{equation} \label{4)} 
\begin{split}
\mathrm{P}{\mathrm{r}}^y_{E_{{\mathrm{L}}_l}\left(y,k\right)}\\&\mathrm{=}\frac{{\left({\mathcal{G}}'^y_{k,B}\mathrm{+}\mathrm{\partial }\right)}^{\chi }}{\sum_{m\mathrm{\in }V\mathrm{-}x}{{\left({\mathcal{G}}'^y_{m,B}\mathrm{+}\mathrm{\partial }\right)}^{\chi }}}\\&\mathrm{=}\frac{{\left(\left({\mathcal{G}}^y_{k,B}e^{\mathrm{-}\tau \left(\mathit{\Delta}B_F\left(E_{{\mathrm{L}}_l}\left(y,k\right)\right)\right)}\mathrm{+}{\lambda }'_{E_{{\mathrm{L}}_l}\left(y,k\right)}\right)\mathrm{+}\mathrm{\partial }\right)}^{\chi }}{\sum_{m\mathrm{\in }V\mathrm{-}x}{{\left({\mathcal{G}}^y_{m,B}e^{\mathrm{-}\tau \left(\mathit{\Delta}B_F\left(E_{{\mathrm{L}}_l}\left(y,m\right)\right)\right)}\mathrm{+}\mathrm{\partial }\right)}^{\chi }}},                
\end{split}
\end{equation} 
while $\mathrm{P}{\mathrm{r}}^y_{E_{{\mathrm{L}}_l}\left(j,y\right)}$ is the probability that link $E_{{\mathrm{L}}_l}\left(j,y\right)$, where $j\mathrm{\in }V\mathrm{-}z$, will be selected at node $y$, expressed as
\begin{equation} \label{5)} 
\begin{split}
\mathrm{P}{\mathrm{r}}^y_{E_{{\mathrm{L}}_l}\left(j,y\right)}\\&\mathrm{=}\frac{{\left({\mathcal{G}}'^y_{A,x}\mathrm{+}\mathrm{\partial }\right)}^{\chi }}{\sum_{g\mathrm{\in }V\mathrm{-}z}{{\left({\mathcal{G}}'^y_{A,g}\mathrm{+}\mathrm{\partial }\right)}^{\chi }}}\\&\mathrm{=}\frac{{\left(\left({\mathcal{G}}^y_{A,x}e^{\mathrm{-}\tau \left(\mathit{\Delta}B_F\left(E_{{\mathrm{L}}_l}\left(j,y\right)\right)\right)}\mathrm{+}{\lambda }'_{E_{{\mathrm{L}}_l}\left(j,y\right)}\right)\mathrm{+}\mathrm{\partial }\right)}^{\chi }}{\sum_{g\mathrm{\in }V\mathrm{-}z}{{\left({\mathcal{G}}^y_{A,g}e^{\mathrm{-}\tau \left(\mathit{\Delta}B_F\left(E_{{\mathrm{L}}_l}\left(g,y\right)\right)\right)}\mathrm{+}\mathrm{\partial }\right)}^{\chi }}}.              
\end{split}
\end{equation} 

\begin{figure*}[h!]
\vspace{-0.6cm}
 \begin{center}
 	 \includegraphics[angle = 0,width=1\linewidth]{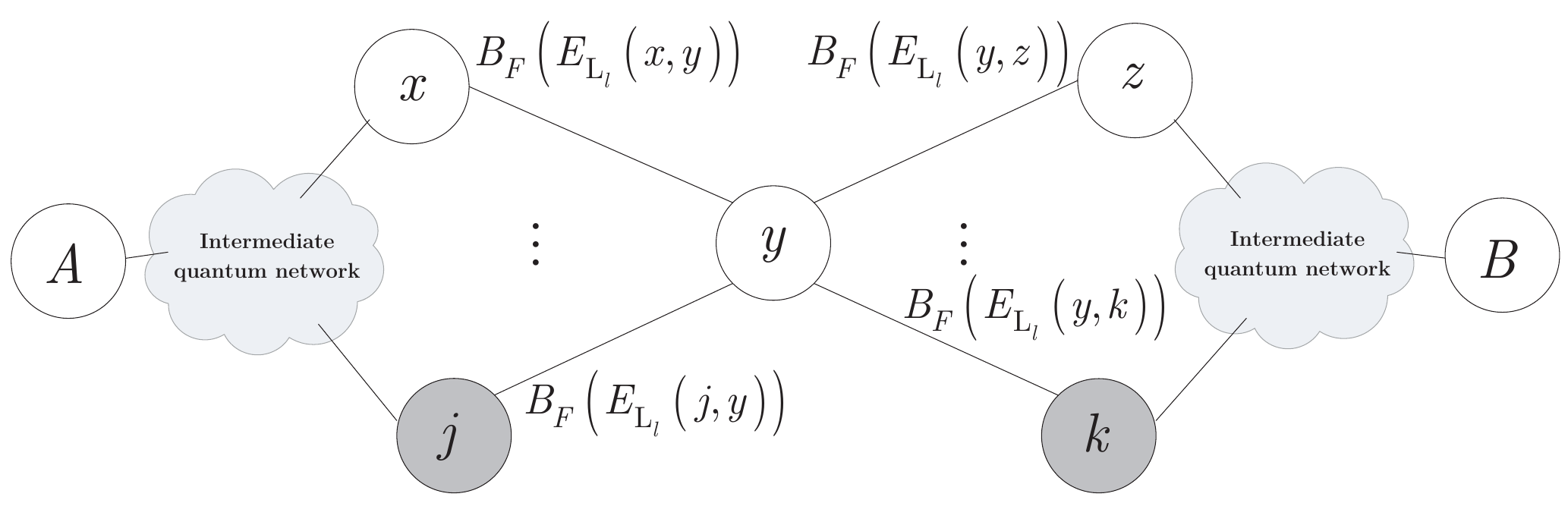}

\caption{Modeling source-dependent link selection through an intermediate network between quantum nodes $A$ and $B$. The current node is $y$, with source node $A$ and destination node $B$. The current direct neighbor of $y$ with a path to source node $A$ is $x$, while node $z$ is the current direct neighbor of $y$ with a path to target node $B$. Node $j$ (gray) models a set of other neighbors of $y$ with paths to source $A$, $j\mathrm{\in }V\mathrm{-}x$, while node $k$ (gray) models a set of other neighbor of $y$ with path to destination $B$, $k\mathrm{\in }V\mathrm{-}z$. Each direct neighbor has en entangled connection with $y$, denoted by $E_{{\mathrm{L}}_l}\left(x,y\right)$, $E_{{\mathrm{L}}_l}\left(y,z\right)$, and $E_{{\mathrm{L}}_l}\left(j,y\right)$, $E_{{\mathrm{L}}_l}\left(y,k\right)$. The entangled links are characterized by entanglement throughputs $B_F\left(E_{{\mathrm{L}}_l}\left(\mathrm{\cdot }\right)\right)$, from which $\mathit{\Delta}B_F\left(E_{{\mathrm{L}}_l}\left(y,z\right)\right)$ is determined in node $y$ to evaluate the entanglement gradient.}
\label{figs1}
\end{center}
\end{figure*}

The model of the intermediate quantum network between $A$ and $B$ used for the derivation of \eqref{ZEqnNum338413} is illustrated in \fref{figs1}. In the source-dependent network model, node $A$ is also a source node for direct neighbor nodes $x$ and $j$ (e.g., exists a path between $A$ and $x$, and between $A$ and $j$), while node $B$ is also a target node for direct neighbor nodes $z$ and $k$  (e.g., exists a path between $z$ and $B$, and between $k$ and $B$ through an intermediate quantum network, respectively). The direct neighbors of $y$ share an entangled link with the current node $y$. 
\\

\newpage
\subsection{Notations}
The notations of the manuscript are summarized in \tref{tab1}.

\begin{longtable}{|p{1.6in}|p{2.8in}|} 
\caption{Summary of notations.}
\label{tab1}
\endfirsthead
\endhead
\hline
\textit{Notation} & \textit{Description} \\ \hline 
L1\textit{} & Manhattan distance (L1 metric). \\ \hline 
$l$  & Level of entanglement.  \\ \hline 
$F$ & Fidelity of entanglement.  \\ \hline 
$N$ & Entangled quantum network, $N\mathrm{=}\left(V,\mathcal{S}\right)$, where $V$ is a set of nodes, $\mathcal{S}$ is a set of entangled links. \\ \hline 
${\mathrm{L}}_l$\textit{} & An $l$-level entangled link. For an ${\mathrm{L}}_l$ link, the hop-distance is ${\mathrm{2}}^{l\mathrm{-}\mathrm{1}}$. \\ \hline 
$d{\left(x,y\right)}_{{\mathrm{L}}_l}$\textit{} & Hop-distance of an $l$-level entangled link between nodes $x$ and $y$, $d{\left(A,B\right)}_{{\mathrm{L}}_l}\mathrm{=}{\mathrm{2}}^{l\mathrm{-}\mathrm{1}}$.  \\ \hline 
$E_{{\mathrm{L}}_l}\left(x,y\right)$ & Entangled link\textit{ }$E_{{\mathrm{L}}_l}\left(x,y\right)$ between nodes $x$ and $y$. \\ \hline 
${\lambda }_{E_{{\mathrm{L}}_l}\left(x,y\right)}$ & Initial entanglement utility of link $E_{{\mathrm{L}}_l}\left(x,y\right)$. \\ \hline 
${\lambda }'_{E_{{\mathrm{L}}_l}\left(x,y\right)}$ & Updated entanglement utility of link $E_{{\mathrm{L}}_l}\left(x,y\right)$. \\ \hline 
$B_F\left(E_{{\mathrm{L}}_l}\left(x,y\right)\right)$ & Entanglement throughput of a given ${\mathrm{L}}_l$-level entangled link $E_{{\mathrm{L}}_l}\left(x,y\right)$ between nodes $\left(x,y\right)$. \\ \hline 
${\mathcal{G}}^y_{A,x}$\textit{} & Initial entanglement gradient, from a source node $A$, on the neighbor node $x$ at $y$, ${\mathcal{G}}^y_{A,x}\mathrm{\ge }\mathrm{0}$. \\ \hline 
\textit{} & Updated entanglement gradient, from source node $A$, on the neighbor node $x$ at $y$, ${\mathcal{G}}^y_{A,x}\mathrm{\ge }\mathrm{0}$. \\ \hline 
$\tau $ & Decay rate of entanglement gradient, $\tau \mathrm{\ge }\mathrm{0}$. \\ \hline 
$f\left(x\right)$ & Probability distribution function. \\ \hline 
$\mathit{\Delta}B_F\left(E_{{\mathrm{L}}_l}\left(x,y\right)\right)$\textit{} & Deviation of entanglement throughput of a given link $E_{{\mathrm{L}}_l}\left(x,y\right)$ from an average, defined as \newline $\mathit{\Delta}B_F\left(E_{{\mathrm{L}}_l}\left(x,y\right)\right) \newline \mathrm{=}\left|\frac{\sum^n_{h\mathrm{=1}}{B_F\left(E_{{\mathrm{L}}_l}\left(y,h\right)\right)}}{n}\mathrm{-}B_F\left(E_{{\mathrm{L}}_l}\left(x,y\right)\right)\right|$,                        \newline where $n$ is the number of direct connections of node $y$, $\sum^n_{h\mathrm{=1}}{B_F\left(E_{{\mathrm{L}}_l}\left(y,h\right)\right)}$ is the total entanglement throughput of all $n$ direct links of node $y$, while $B_F\left(E_{{\mathrm{L}}_l}\left(x,y\right)\right)$ is the entanglement throughput of link $E_{{\mathrm{L}}_l}\left(x,y\right)$ between nodes $y$ and $x$. \\ \hline 
$e^x$ & Exponential distribution function. \\ \hline 
$X^y_{E_{{\mathrm{L}}_l}\left(x,y\right)}\left(t\right)$ & Non-negative, non-stationary random process of entanglement utility ${\lambda }_{E_{{\mathrm{L}}_l}\left(x,y\right)}$. \\ \hline 
${\mu }^y_{E_{{\mathrm{L}}_l}\left(x,y\right)}\left(t\right)$ & Mean of a non-negative, non-stationary random process $X^y_{E_{{\mathrm{L}}_l}\left(x,y\right)}\left(t\right)$. \\ \hline 
$E\left[X^y_{E_{{\mathrm{L}}_l}\left(x,y\right)}\left(t\right)\right]$ & Estimate of $X^y_{E_{{\mathrm{L}}_l}\left(x,y\right)}\left(t\right)$, defined as\newline $E\left[X^y_{E_{{\mathrm{L}}_l}\left(x,y\right)}\left(t\right)\right]\mathrm{=}X^y_{E_{{\mathrm{L}}_l}\left(x,y\right)}\left(t\right)\mathrm{*}{\mathrm{\Omega }}_{{\mathcal{G}}^y_{A,x}}\left(t\right)$,\newline where $\mathrm{*}$ is the convolution operator, while function ${\mathrm{\Omega }}_{{\mathcal{G}}^y_{A,x}}\left(t\right)$ is defined as\newline ${\mathrm{\Omega }}_{{\mathcal{G}}^y_{A,x}}\left(t\right)\mathrm{=}e^{\mathrm{-}\tau \left(\mathit{\Delta}B_F\left(E_{{\mathrm{L}}_l}\left(x,y\right)\right)\right)}U\left(t\right)$,\newline where $U\left(t\right)$is the unit step function.  \\ \hline 
${\mathrm{\bot }}_{{\mathcal{G}}^y_{A,x}}\left(\mathrm{\Delta }T\right)$ & Correlation function, ${\mathrm{\bot }}_{{\mathcal{G}}^y_{A,x}}\left(\mathrm{\Delta }T\right)\mathrm{=}e^{\mathrm{-}\tau \left|\mathrm{\Delta }T\right|}$, where $\mathrm{\Delta }T$ is a time period. \\ \hline 
$\mathrm{P}{\mathrm{r}}^y_{E_{{\mathrm{L}}_l}\left(y,z\right)}$ & Link selection probability, probability that from node $y$ the entangled link $E_{{\mathrm{L}}_l}\left(y,z\right)$ is selected to reach destination $B$.  \\ \hline 
$\mathrm{\partial }$ & Threshold parameter, $\mathrm{\partial }\mathrm{\ge }\mathrm{0}$. \\ \hline 
$\chi $\textit{} & Tuning parameter, $\chi \mathrm{\ge }\mathrm{0}$. \\ \hline 
${\mathcal{P}}_i$ & An $i$-th path between a source node $A$ and target node $B$. \\ \hline 
${\mathcal{G}}^A_{{\mathcal{P}}_i}$ & Initial path entanglement gradient of a given entangled path ${\mathcal{P}}_i$, $i\mathrm{=1,\dots ,}m$ at source node $A$. \\ \hline 
${\mathcal{G}}^B_{{\mathcal{P}}_i}$ & Initial path entanglement gradient of ${\mathcal{P}}_i$, $i\mathrm{=1,\dots ,}m$ at destination node $B$. \\ \hline 
${\kappa }_A$\textit{} & Observation rate, mean number of $d$-dimensional entangled states arrive in $A$. \\ \hline 
${\kappa }_B$\textit{} & Observation rate, mean number of $d$-dimensional entangled states arrive in $B$. \\ \hline 
${\kappa }_{AB}$ & Total observation rate, for a symmetrical arrival of the entangled states, ${\kappa }_A\mathrm{=}{\kappa }_B\mathrm{=}{{\kappa }_{AB}}/{\mathrm{2}}$. \\ \hline 
$\mathrm{Z}$ & Random variable, $Z\mathrm{=}e^{\mathrm{-}K\tau }$ where $K$ is a random variable which models the interarrival time between the entangled states. \\ \hline 
$\mu \left(K\right)$ & Mean of random variable $K$. \\ \hline 
$f_Z\left(x\right)$ & Probability distribution function of $Z$, \newline $f_Z\left(x\right)\mathrm{=}\frac{{\kappa }_{AB}}{\tau }x^{\left(\frac{{\kappa }_{AB}}{\tau }\mathrm{-}\mathrm{1}\right)}$,\newline where $\mathrm{0}\mathrm{\le }x\mathrm{\le }\mathrm{1}$.     \\ \hline 
$\mu \left(Z\right)$ & Mean of random variable $Z\mathrm{=}e^{\mathrm{-}K\tau }$,\newline   $\mu \left(Z\right)\mathrm{=}\frac{{\kappa }_{AB}}{{\kappa }_{AB}\mathrm{+}\tau }\mathrm{=}{\gamma }_{AB}$. \\ \hline 
${\mathcal{G}}'^A_{{\mathcal{P}}_i}$ & Updated path entanglement gradient the source node $A$ for a given path ${\mathcal{P}}_i$, $i\mathrm{=1,\dots ,}m$. \\ \hline 
${\mathcal{G}}'^B_{{\mathcal{P}}_i}$ & Updated path entanglement gradient the source node $B$ for a given path ${\mathcal{P}}_i$, $i\mathrm{=1,\dots ,}m$. \\ \hline 
${\mathcal{G}}'^A_{{\mathcal{P}}_j}$ & Updated path entanglement gradient the source node $A$ for a given path ${\mathcal{P}}_j$, $j\mathrm{\ne }i$. \\ \hline 
${\mathcal{G}}'^B_{{\mathcal{P}}_j}$ & Updated path entanglement gradient the source node $B$ for a given path ${\mathcal{P}}_j$, $j\mathrm{\ne }i$. \\ \hline 
${\mu }^A_{{\mathcal{P}}_i}$ & Average value of received entanglement gradient from path ${\mathcal{P}}_i$, $i\mathrm{=1,\dots ,}m$at node $A$. \\ \hline 
${\mu }^B_{{\mathcal{P}}_i}$ & Average value of received entanglement gradient from path ${\mathcal{P}}_i$, $i\mathrm{=1,\dots ,}m$at node $B$. \\ \hline 
${\mu }^A_{{\mathcal{P}}_j}$ & Average value of received entanglement gradient from path ${\mathcal{P}}_i$, $j\mathrm{\ne }i$, at node $A$. \\ \hline 
${\mu }^B_{{\mathcal{P}}_j}$ & Average value of received entanglement gradient from path ${\mathcal{P}}_i$, $j\mathrm{\ne }i$, at node $B$. \\ \hline 
${\mathcal{P}}^{\mathrm{*}}$ & Optimal shortest path. \\ \hline 
${\mathcal{G}}'^A_{{\mathcal{P}}^{\mathrm{*}}}$ & Updated path entanglement gradient the source node $A$ for optimal shortest path ${\mathcal{P}}^{\mathrm{*}}$. \\ \hline 
$\mathrm{P}{\mathrm{r}}^A_{{\mathcal{P}}_i}$ & Probability that path ${\mathcal{P}}_i$, $i\mathrm{=1,\dots ,}m$ will be used by node $A$.  \\ \hline 
$\mathrm{P}{\mathrm{r}}^B_{{\mathcal{P}}_i}$ & Probability that path ${\mathcal{P}}_i$, $i\mathrm{=1,\dots ,}m$ will be used by node $B$. \\ \hline 
$\mathrm{P}{\mathrm{r}}^A_{{\mathcal{P}}_j}$ & Probability that path ${\mathcal{P}}_j$, $j\mathrm{\ne }i$, will be used by node $A$. \\ \hline 
$\mathrm{P}{\mathrm{r}}^B_{{\mathcal{P}}_j}$ & Probability that path ${\mathcal{P}}_j$, $j\mathrm{\ne }i$, will be used by node $B$. \\ \hline 
$\mathbb{E}\left({\mathcal{G}}'^A_{{\mathcal{P}}_i}\right)$ & Mean entanglement gradient of a particular path ${\mathcal{P}}_i$ at $A$, $i\mathrm{=1,\dots ,}m$. \\ \hline 
$\mathbb{E}\left({\mathcal{G}}'^B_{{\mathcal{P}}_i}\right)$ & Mean entanglement gradient of a particular path ${\mathcal{P}}_i$ at $B$, $i\mathrm{=1,\dots ,}m$. \\ \hline 
${\tau }_{\mathbb{E}\left({\mathcal{G}}'^n_{{\mathcal{P}}_i}\right)}$ & Decay rate of mean path entanglement gradient $\mathbb{E}\left({\mathcal{G}}'^n_{{\mathcal{P}}_i}\right)$. \\ \hline 
${\mathrm{\partial }}_{\mathbb{E}\left({\mathcal{G}}'^n_{{\mathcal{P}}_i}\right)}$ & Threshold parameter to yield the ${\tau }_{\mathbb{E}\left({\mathcal{G}}'^n_{{\mathcal{P}}_i}\right)}$ decay rate of mean path entanglement gradient. \\ \hline 
${\widetilde{\tau }}_{\mathbb{E}\left({\mathcal{G}}'^n_{{\mathcal{P}}_i}\right)}$ & Optimal estimator of ${\tau }_{{\mathcal{G}}'^n_{{\mathcal{P}}_i}}$. \\ \hline 
$Y$ & Variable. \\ \hline 
$B_F\left({\mathcal{P}}_i\right)$ & Entanglement throughput (measured in $d$-dimensional entangled states of a particular fidelity $F$ per sec) of path ${\mathcal{P}}_i$ \\ \hline 
${\tilde{B}}_F\left({\mathcal{P}}_i\right)$ & An expected ${\tilde{B}}_F\left({\mathcal{P}}_i\right)$ entanglement throughput of a path ${\mathcal{P}}_i$. \\ \hline 
$\varphi \left({\mathcal{P}}_i\right)$ & Deviation of a current $B_F\left({\mathcal{P}}_i\right)$ entanglement throughput (measured in $d$-dimensional entangled states of a particular fidelity $F$ per sec) of path ${\mathcal{P}}_i$ from an expected ${\tilde{B}}_F\left({\mathcal{P}}_i\right)$ entanglement throughput of path ${\mathcal{P}}_i$, as\newline $\varphi \left({\mathcal{P}}_i\right)\mathrm{=}\left|{\tilde{B}}_F\left({\mathcal{P}}_i\right)\mathrm{-}B_F\left({\mathcal{P}}_i\right)\right|$. \\ \hline 
${\mathrm{\Phi }}^{s,n}_{{\mathcal{P}}_i}$ & Parameter for a given path ${\mathcal{P}}_i$, between a source node $s$ and current node $n$, defined as \newline ${\mathrm{\Phi }}^{s,n}_{{\mathcal{P}}_i}\mathrm{=}\sum^n_{x\mathrm{=}s}{\alpha {\sigma }^x_{{\mathcal{P}}_i}}$,\newline where $\alpha $ and ${\sigma }^x_{{\mathcal{P}}_i}$ are coefficients. \\ \hline 
${\sigma }^x_{{\mathcal{P}}_i}$ & Coefficient used by ${\mathrm{\Phi }}^{s,n}_{{\mathcal{P}}_i}$, \newline ${\sigma }^x_{{\mathcal{P}}_i}\mathrm{=log}\left(\frac{{\mathcal{G}}'^{x\mathrm{+1}\mathrm{\in }{\mathcal{P}}_i}_{{\mathcal{P}}_i}}{{\mathcal{G}}'^{x\mathrm{\in }{\mathcal{P}}_i}_{{\mathcal{P}}_i}}\right)$,\newline where ${\mathcal{G}}'^{x\mathrm{\in }{\mathcal{P}}_i}_{{\mathcal{P}}_k}$ is the entanglement gradient of node $x\mathrm{\in }{\mathcal{P}}_i$, while ${\mathcal{G}}'^{x\mathrm{+1}\mathrm{\in }{\mathcal{P}}_i}_{{\mathcal{P}}_k}$ is the entanglement gradient at node $x\mathrm{+1}\mathrm{\in }{\mathcal{P}}_i$. \\ \hline 
$\alpha $ & Coefficient used by ${\mathrm{\Phi }}^{s,n}_{{\mathcal{P}}_i}$, defined as\newline $\alpha \mathrm{=}\left\{ \begin{array}{l}
\mathrm{1,}\textnormal{ if }\left|{\sigma }^x_{{\mathcal{P}}_i}\right|\mathrm{>}\vartheta  \\ 
0,\textnormal{ if }\left|{\sigma }^x_{{\mathcal{P}}_i}\right|\mathrm{\le }\vartheta  \end{array}
\right.$,\newline where $\vartheta $ is a threshold. \\ \hline 
$\vartheta $ & Threshold parameter. \\ \hline 
${\mu }^n\left({\mathrm{\Phi }}^{s,n}_{\mathcal{P}}\right)$ & Mean for the $m$ paths ${\mathcal{P}}_{\mathrm{1}}\mathrm{,\dots ,}{\mathcal{P}}_m$ between a source node $s$ and a current node $n$, as\newline ${\mu }^n\left({\mathrm{\Phi }}^{s,n}_{\mathcal{P}}\right)\mathrm{=}\frac{\sum^m_{i\mathrm{=1}}{{\mathrm{\Phi }}^{s,n}_{{\mathcal{P}}_i}}}{m}$. \\ \hline 
$\psi \left(n,z\right)$ & A distance function $\psi \left(n,z\right)$ between $n$ and $z$. \\ \hline 
$\mathbb{E}\left({\mathcal{G}}'^n_{{\mathcal{P}}_i}\right)$ & Mean entanglement gradients at node $n\mathrm{\in }{\mathcal{P}}_i$. \\ \hline 
$\mathbb{E}\left({\mathcal{G}}'^z_{{\mathcal{P}}_i}\right)$ & Mean entanglement gradients at node $z\mathrm{\in }{\mathcal{P}}_i$. \\ \hline 
${\theta }^z_{E_{{\mathrm{L}}_l}\left(n,z\right)}$ & Inverse link entanglement gradient, \newline ${\theta }^z_{E_{{\mathrm{L}}_l}\left(n,z\right)}\newline\mathrm{=}\frac{\mathrm{1}}{{\mathcal{G}}'^z_{A,n}}\mathrm{=}\frac{\mathrm{1}}{{\mathcal{G}}^z_{A,n}e^{\mathrm{-}\tau \left(\mathit{\Delta}B_F\left(E_{{\mathrm{L}}_l}\left(n,z\right)\right)\right)}\mathrm{+}{\lambda }'_{E_{{\mathrm{L}}_l}\left(n,z\right)}}$,                         \newline where ${\lambda }'_{E_{{\mathrm{L}}_l}\left(n,z\right)}$ is the updated entanglement utility, as\newline ${\lambda }'_{E_{{\mathrm{L}}_l}\left(n,z\right)}\mathrm{=}\frac{{\lambda }_{E_{{\mathrm{L}}_l}\left(n,z\right)}}{\mathrm{1+}B_F\left(E_{{\mathrm{L}}_l}\left(n,z\right)\right){\lambda }_{E_{{\mathrm{L}}_l}\left(n,z\right)}}.$ \\ \hline 
$t$  & Number of threads. \\ \hline 
${\mathcal{T}}_i$ & An $i$-th thread, ${\mathcal{T}}_{\mathrm{1}}\mathrm{,\dots ,}{\mathcal{T}}_t$. \\ \hline 
${\ell }_{\mathcal{T}}$ & Thread-threshold, limits maximal number of nodes visited by a given thread to at most ${\ell }_{\mathcal{T}}$. \\ \hline 
$p_{{\mathcal{T}}_i}\left(n,z\right)$ & Link selection probability for an $i$-th thread ${\mathcal{T}}_i$. \\ \hline 
${\mathrm{S}}_{{\mathcal{T}}_i}$ & A set of nodes already visited by the $i$-th thread ${\mathcal{T}}_i$. \\ \hline 
$\mathrm{P}{\mathrm{r}}_{{\mathcal{T}}_i}\left(n,z\right)$ & Probability function for an $i$-th thread ${\mathcal{T}}_i$. \\ \hline 
$C_{\mathrm{1}}$, $C_{\mathrm{2}}$ & Weighting parameters to balance the relevance between inverse entanglement gradient function $\theta \left(\mathrm{\cdot }\right)$ and distance function $\psi \left(\mathrm{\cdot }\right)$ in $\mathrm{P}{\mathrm{r}}_{{\mathcal{T}}_i}\left(n,z\right)$. \\ \hline 
${\nu }_n$, $\varsigma \left({\gamma }_n\right)$, $\rho \left({\nu }_n\right)$ & Additional parameters. \\ \hline 
${\gamma }_n$ & Parameter for a node $n$, defined as \newline ${\gamma }_n\mathrm{=}\frac{{\kappa }_n}{{\kappa }_n\mathrm{+}{\tau }_n}\mathrm{=}{\left(\mathrm{1+}\frac{{\tau }_n}{{\kappa }_n}\right)}^{\mathrm{-}\mathrm{1}}$,\newline where ${\kappa }_n$ is the observation rate in node $n$, mean number of $d$-dimensional entangled states arrive in $n$, ${\tau }_n$ decay rate of entanglement gradient in node $n$.  \\ \hline 
$\mathrm{\Pi }$ & Tuning parameter (a fraction of peak value), $\mathrm{0}\mathrm{\le }\mathrm{\Pi }\mathrm{\le }\mathrm{1}$. \\ \hline 
${\kappa }^{\mathrm{*}}_n$ & Cutoff observation rate (critical value of received $d$-dimensional entangled states per sec) defined at a given observation rate ${\kappa }_n$, controllable by ${\tau }_n$. \\ \hline 
$\mathcal{R}$ & Ideal recovery operation with an optimal quantum error correction. \\ \hline 
$\widetilde{\mathrm{\Psi }}$ & Shared Bell pair between the final stations. \\ \hline 
${\rho }_f$ & Input density matrix of ideal recovery operation $\mathcal{R}$. \\ \hline 
$P_{err}$ & Per-node error probability $P_{err}$, includes the effective logical error probability $Q$ and other residual errors ${\varepsilon }_{res}$ in the node. \\ \hline 
$\mathcal{M}\left(A,B\right)$ & Correlation measurement between the final stations $A$ and $B$, yields entanglement fidelity as $\mathcal{M}\left(A,B\right)\mathrm{\approx }\sqrt{F}.$ \\ \hline 
\end{longtable}
\end{document}